\documentclass[final,5p,times,twocolumn]{elsarticle}

\usepackage{xurl}
\usepackage[utf8]{inputenc}
\usepackage[T1]{fontenc}
\usepackage{amsmath,amssymb}
\usepackage{graphicx}
\usepackage{booktabs}
\usepackage{multirow}
\usepackage{array}
\usepackage{tabularx}
\usepackage{longtable}
\usepackage{hyperref}
\usepackage{xcolor}
\usepackage{enumitem}
\usepackage{caption}
\usepackage[most]{tcolorbox}
\usepackage[framemethod=default]{mdframed}
\usepackage{needspace}
\usepackage{colortbl}
\usepackage{makecell}
\usepackage{microtype}

\newcolumntype{L}[1]{>{\raggedright\arraybackslash}p{#1}}
\newcolumntype{S}{>{\centering\arraybackslash}p{0.75cm}}

\begin{document}

\begin{frontmatter}

\title{Governing What the EU AI Act Excludes: Accountability for Autonomous AI Agents in Smart City Critical Infrastructure}

\author[hct]{Talal Ashraf Butt\corref{cor1}}
\ead{tbutt@hct.ac.ae}
\author[hct]{Muhammad Iqbal}
\ead{miqbal1@hct.ac.ae}
\author[cmu]{Razi Iqbal}
\ead{razi.iqbal@ieee.org}
\cortext[cor1]{Corresponding author}
\affiliation[hct]{organization={Higher Colleges of Technology},
            city={Fujairah},
            country={United Arab Emirates}}
\affiliation[cmu]{organization={Central Michigan University},
            city={Mount Pleasant},
            country={United States of America}}

\begin{abstract}
When a traffic signal controller adjusts green phases and a grid manager curtails power on the same corridor, each system may comply with its own obligations. The resident who suffers the combined effect has no single authority to hold accountable and, under the EU AI Act, limited means to obtain an explanation. Annex~III, point~2 excludes safety-component AI in critical infrastructure from Article~86 explanation rights and Article~27 fundamental-rights impact assessment. Provider and deployer duties under Articles~9--15 still apply, and residual pathways under the GDPR, NIS2, and tortious liability offer partial coverage. The Act's principal resident-facing accountability instruments are nonetheless narrowed for the autonomous infrastructure systems most likely to interact across agencies.

The paper's regulatory analysis traces this accountability deficit through four residual pathways (GDPR Article~22, GDPR transparency obligations, tortious liability, and NIS2) and shows that each is structurally bounded by individual-controller, individual-decision scope. As a governance response, the paper presents AgentGov-SC, a three-layer architecture (Agent, Orchestration, City) specifying 25 governance measures with bidirectional traceability to the EU AI Act, ISO/IEC~42001, and the NIST AI Risk Management Framework. Five cross-framework conflict resolution rules and an autonomy-calibrated activation model complete the design. A structured scenario analysis traces governance activation through a multi-agent corridor cascade involving three documented UAE smart-city systems, with a contrasting single-system scenario confirming proportional activation. As the August~2026 full-applicability deadline approaches, the paper contributes a regulatory gap analysis and governance architecture for an increasingly important class of urban AI deployment that existing frameworks treat as bounded and isolated.
\end{abstract}


\end{frontmatter}

\section{Introduction}
\label{sec:introduction}

The EU AI Act narrows its principal resident-facing accountability pathways for some of the AI systems most deeply embedded in urban life. Article~86(1) grants affected persons a right to obtain clear and meaningful explanations of decisions produced by high-risk AI systems listed in Annex~III, but carves out systems classified under point~2 of that Annex~\citep{euaiact_2024}. Traffic signal controllers, grid management systems, and surveillance platforms are point~2 systems, and they are the systems most likely to interact across agencies in a single urban corridor. Article~27(1) imposes the Fundamental Rights Impact Assessment on public-body deployers of high-risk AI systems across all Annex~III use cases except point~2~\citep{euaiact_2024,ec_aiact_servicedesk_2025_art27}. Article~49 further limits EU-level database registration to systems under those same Annex~III points. Point~2 systems fall under national registry arrangements instead~\citep{euaiact_2024}. Annex~III, point~2 covers AI intended to be used as safety components in the management and operation of critical digital infrastructure, road traffic, or the supply of water, gas, heating, or electricity~\citep{euaiact_2024}. In combination, explanation rights, structured fundamental-rights impact assessment, and centralised registration are narrowed for safety-component AI in core smart-city infrastructure domains.

Three qualifications constrain this observation. First, the carve-out applies to safety components under Article~3(14). Energy distribution controllers and traffic signal systems qualify straightforwardly. Surveillance systems present a harder case, falling more plausibly under Annex~III, point~5 (law enforcement), where Articles~86 and~27 apply without exemption~\citep{kaminski_malgieri_2025_explanation}. Even so, the interaction-level gap persists: no single-system explanation right reconstructs a causal chain spanning both exempt and non-exempt systems. Second, provider and deployer duties under Articles~9--15 remain applicable~\citep{euaiact_2024}. The gap concentrates in resident-facing remedies. Third, GDPR, NIS2, and tortious liability offer partial coverage that Section~\ref{sec:bg-regulatory} examines in detail. The result is not an absence of obligations. It is a narrower set of resident-facing accountability pathways where infrastructure autonomy carries the highest consequence. Section~\ref{sec:bg-regulatory} shows that the sectoral instruments Recital~55 contemplates provide operational continuity safeguards but not the resident-facing accountability instruments that Articles~86 and~27 were designed to supply.

That narrowing becomes structurally significant once the governed object shifts. A bounded AI system can be individually assessed and explained. A set of interacting autonomous agents cannot, because consequences emerge from coordination that no single assessment captures. Agents plan and execute multi-step actions. They invoke tools and delegate subtasks to other agents. When conditions change, they adapt, and that adaptation is the governance problem. A system that modifies its own strategy at runtime creates risks that no static compliance check can see~\citep{shavit_2023_practices,chan_2024_visibility}. In smart cities, this interaction surface is dense. Traffic signal coordination and energy distribution share physical infrastructure across separate authorities. Surveillance enforcement draws on the same sensor feeds, adding a third agency to an already fragmented governance space. Small coordination choices can produce city-scale effects that no single-system audit would detect. A December~2025 substation fire in San Francisco knocked out traffic signals, and Waymo's autonomous robotaxis stopped in place, blocking arterial routes needed by emergency vehicles~\citep{techcrunch_2025_waymo_sf_blackout}. The event was localised. The structural vulnerability it exposed is not. The governance challenge is not only that autonomous systems make consequential decisions. It is that consequences emerge through interaction, while accountability instruments, including those in the EU AI Act, remain structured largely around individual systems.

Practitioner and institutional work has recognised parts of this challenge without resolving the combination. Proportional safeguards are now a stated priority~\citep{wef_2025_ai_agents}, and Singapore's January~2026 governance framework for agentic AI marks the first national-level attempt at agent-specific guidance~\citep{imda_2026_agentic_mgf}. That framework is scoped to organisational deployment. It does not address the multi-authority conditions that define smart-city infrastructure: three agencies sharing a corridor, each running autonomous systems that affect the others' operations. Security-focused work maps a different surface. OWASP's Top~10 for Agentic Applications identifies risks tied to tool use and delegation~\citep{owasp_2025_agentic_top10}, while policy research proposes dimensional governance built on continuously tracked properties~\citep{engin_hand_2025_dimensional} and oversight structures matched to public-sector operational tempo~\citep{schmitz_2025_oversight_public_sector}. No existing contribution traces governance controls to regulatory obligations, enforces constraints at runtime as agents act, and provides residents with contestation pathways when automated decisions go wrong. That integration is what AgentGov-SC attempts.

Current public-sector AI governance rests on three instruments~\citep{euaiact_2024,isoiec_42001_2023,nist_2023_airmf} that between them cover value-chain obligations, organisational management, and risk-based decision support. All three supply usable baselines, and all three presume that the governed object is a bounded system. The Act does not specify how oversight duties distribute when a traffic signal vendor, a city procurement agency, and a maintenance contractor share operational responsibility for the same deployment~\citep{veale_borgesius_2021_demystifying}. In a corridor like Dubai's E11, behaviour emerges from coordination across agents, shared sensor feeds, and vendor-operated subsystems that no individual deployer can see end to end. Section~\ref{sec:compliance} traces the bounded-system assumption through all three instruments.

Existing governance proposals cover parts of this space without integrating them. MI9 provides runtime enforcement without regulatory traceability~\citep{wang_2025_mi9}. UCF demonstrates cross-regime compliance mapping without agent-specific controls~\citep{eisenberg_2025_ucf}. The recurring pattern is fragmentation: runtime enforcement, compliance mapping, and resident-facing accountability each appear in separate proposals, never in a single architecture designed for multi-agent, real-time operation. Section~\ref{sec:background} traces this fragmentation and positions AgentGov-SC against it.

No meaningful channel exists for residents to contest automated decisions that shape daily routines~\citep{green_2019_smart_enough_city,cardullo_kitchin_2019_citizenship}. Most cannot discover that such a decision was made in the first place. Consider a resident fined by a camera system. The fine can be challenged. What cannot be discovered is what algorithm selected that intersection for enforcement, or why the detection threshold was set where it was. Accountability depends on what agencies choose to document, not on enforceable disclosure obligations~\citep{brauneis_goodman_2018_transparency}. Stated principles do not close this gap. Enforceable governance mechanisms remain the central implementation challenge~\citep{cobbe_2021_reviewable,mittelstadt_2019_principles}. Power and discrimination are structural features of this setting~\citep{crawford_2021_atlas,benjamin_2019_race}, and governance design must treat affected communities as epistemic authorities~\citep{birhane_2021_algorithmic_injustice}. These concerns are not incidental to the regulatory analysis above; they explain why the Annex~III, point~2 carve-out matters. The residents most exposed to infrastructure automation, including those in expatriate-majority urban corridors, transit-dependent communities, and labour housing districts, may also face the greatest difficulty in navigating multi-agency accountability chains when automated decisions go wrong.

This paper presents AgentGov-SC, a three-layer governance architecture for agentic AI in smart-city critical infrastructure, designed as a regulatory response to the accountability deficit identified above.

\noindent\textbf{Contributions.}

\begin{enumerate}[leftmargin=*, label=\textbf{C\arabic*:}]
    \item \textbf{Regulatory gap analysis.} Identification of a compound narrowing of the EU AI Act's resident-facing accountability pathways for autonomous AI in smart-city critical infrastructure. The Annex~III, point~2 exclusion removes Articles~86, 27, and~49 safeguards for safety-component AI. Residual pathways under the GDPR, NIS2, and tortious liability are assessed and shown to be structurally bounded by individual-controller, individual-decision scope.
    \item \textbf{Governance architecture as regulatory response.} AgentGov-SC, a three-layer architecture (Agent, Orchestration, City) specifying 25 governance measures with bidirectional traceability to three regulatory and standards instruments~\citep{euaiact_2024,isoiec_42001_2023,nist_2023_airmf}. The architecture operationalises Article~14 human oversight for multi-agent settings, introduces five cross-framework conflict resolution rules, and calibrates governance intensity to societal impact. A structured scenario analysis traces governance activation through a multi-agent corridor cascade involving three documented UAE smart-city systems, with a contrasting single-system scenario confirming proportional activation.
\end{enumerate}

\section{Regulatory and Governance Background}
\label{sec:background}

This section traces the accountability deficit through the EU AI Act's own provisions, assesses four residual pathways that partially compensate for it, and maps the institutional and scholarly responses that inform AgentGov-SC's design.

\subsection{Regulatory Gap: The EU AI Act and Agentic Autonomy}
\label{sec:bg-regulatory}
 
As established in Section~\ref{sec:introduction}, the EU AI Act's accountability architecture rests on a bounded-system assumption. Article~14 requires human oversight for high-risk systems. How that oversight should work when authority is distributed across interacting agents remains unspecified~\citep{euaiact_2024}. Within a smart-city control environment, oversight becomes an orchestration property. It depends on who can pause an agent and who can override decisions. It depends equally on what trace evidence is retained and whether that evidence crosses vendor boundaries. Accountability breaks down when operational responsibility is split across contractors, agencies, and their automated delegates. Unit-level compliance cannot bridge that gap.
 
The resident-facing gap identified in Section~\ref{sec:introduction} follows from a design choice in the Act. It reflects the Act's underlying assumption that high-risk AI systems can be individually assessed and individually explained~\citep{hacker_2023_ai_liability,stuurman_2022_labelling}. That assumption holds for many deployments. It breaks when the governed object is the interaction space between systems. Meaningful explanation requires causal attribution~\citep{wachter_2018_counterfactual}, which in multi-agent settings demands cross-system traceability that no single deployer can provide. The Annex~III, point~2 exclusion narrows Article~86 relative to GDPR Article~22 in ways that \citet{kaminski_malgieri_2025_explanation} examine in detail. The comparison deserves closer scrutiny. Article~22 provides a decision-specific remedy: it enables a data subject to contest a particular automated decision and to obtain human intervention by the controller responsible for that decision. In multi-agent infrastructure settings, it encounters three structural limitations. The controller receiving an Article~22 challenge may have no access to the decision logs of the other systems whose actions materially shaped the conditions that produced the contested outcome. Article~22 imposes no obligation on those other controllers to disclose their decision logic or contribute to cross-system causal reconstruction. And the ``solely automated'' scope condition may not be met where control-room operators nominally supervise each system, even if no operator saw the interaction that produced the harm. Article~86, by contrast, was designed to require a system-level explanation of how the AI contributed to the decision. It is precisely this broader instrument that the Annex~III, point~2 carve-out withdraws. The gap is therefore not an absence of any remedy. It is the absence of a remedy commensurate with the cross-system causal complexity that agentic infrastructure produces. Laux, Wachter, and Mittelstadt~\citep{laux_2024_trustworthy} argue more broadly that the Act conflates trustworthiness with risk acceptability, which creates implementation challenges precisely where risk is distributed across institutional boundaries.

Recital~55 explains the legislative rationale. The carve-out reflects an expectation that sectoral Union legislation already provides equivalent safeguards for critical infrastructure AI~\citep{euaiact_2024}. The expectation is sound as a principle of regulatory coherence. But which sectoral instruments does it contemplate? Sectoral legislation protects operational reliability in energy and road traffic.\footnote{Electricity Market Directive~(EU)~2019/943, Critical Entities Resilience Directive~(EU)~2022/2557, and Directive~2010/40/EU on Intelligent Transport Systems.} None of these instruments creates a resident-facing right to an explanation of how an automated infrastructure decision was taken. NIS2 adds cybersecurity risk management and incident reporting across both sectors. None of these instruments creates a resident-facing right to an explanation of how an automated infrastructure decision was taken. None requires cross-system causal reconstruction when multiple autonomous systems interact to produce a harmful outcome. The equivalent-safeguards assumption therefore holds for operational continuity and system-level safety. It does not hold for the specific accountability instruments that Articles~86 and~27 were designed to provide: structured, individual-facing remedies that enable affected persons to understand and contest decisions based on AI outputs. The analysis below traces this gap through four residual pathways.

GDPR transparency and impact assessment obligations reach this processing but stop at the individual controller~\citep{gdpr_2016_679}.\footnote{Specifically Articles~13(2)(f), 14(2)(g), 15(1)(h), and~35.} No provision requires one controller to disclose how its automated decisions interact with another's. DPIAs are controller-specific, however. Each agency assesses its own processing, with no provision for joint assessment across controllers whose systems interact to produce cumulative effects. Recent CJEU case law has strengthened individual-system transparency. In \textit{SCHUFA}, the Court confirmed that automated credit scoring constitutes a decision under Article~22~\citep{cjeu_2023_schufa}. The February~2025 \textit{Dun \& Bradstreet Austria} judgment went further, holding that meaningful information about the logic involved cannot be reduced to vague descriptions~\citep{cjeu_2025_dun_bradstreet}. Both rulings reinforce accountability for individual automated decisions. Cross-system interaction remains outside their reach. The GDPR's residual layer is therefore more substantial than a narrow reading of Article~22 alone suggests. It remains, however, structurally bounded by individual-controller, individual-decision scope.

Tortious liability pathways have shifted since the AI Act was adopted. The proposed AI Liability Directive (COM(2022)~496~final) was withdrawn in February~2025~\citep{ec_2025_aild_withdrawal}. It would have introduced rebuttable presumptions of causality and structured evidence disclosure for AI-related harm. Affected persons must instead rely on the revised Product Liability Directive~(EU)~2024/2853. That instrument extends strict liability to software and AI-enabled products and allows courts to order disclosure of technical documentation~\citep{pld_2024_2853}. For multi-agent cascades, however, the PLD operates ex~post through litigation. It does not provide the ex~ante, structured contestation pathways that Articles~86 and~27 were designed to supply. Establishing causation across three autonomous systems operated by separate authorities under separate contracts imposes an evidentiary burden that the withdrawn Directive's mechanisms were specifically designed to alleviate. The gap therefore survives the current liability framework, albeit in narrower form.

NIS2 adds a further governance layer. Smart-city infrastructure operators may qualify as essential entities, subjecting them to cybersecurity risk management and incident-handling obligations~\citep{nis2_2022_2555}. For incident response, this creates the multi-clock tension addressed by AgentGov-SC's T4 resolution rule (Table~\ref{tab:conflicts}). Beyond reporting timelines, however, NIS2's risk-management obligations (Article~21) operate at the entity level. They do not require assessment of cross-entity interaction risks between autonomous systems operated by different essential entities sharing urban infrastructure. The gap between entity-level cybersecurity governance and system-of-systems interaction governance is structurally analogous to the AI Act's bounded-system assumption.

Harmonised standards may narrow the gap over time. CEN-CENELEC JTC~21 is developing standards on risk management and transparency, with additional work on human oversight and conformity assessment~\citep{ec_2024_ai_act_standardisation}. Once referenced in the Official Journal, these create a presumption of conformity under Article~40. Whether standardisation will address cross-system interaction remains uncertain, since the standards are being developed around individual AI system compliance. The deficit described in this paper is therefore currently operative, though legally contingent. Future standards and delegated acts may narrow it if standardisation efforts extend to interaction-level risk management~\citep{ec_2025_ai_system_definition,ec_2025_prohibited_practices}.
The Digital Omnibus proposed by the European Commission in November~2025~\citep{ec_2025_digital_omnibus} originally linked the applicability of high-risk AI system obligations to the availability of harmonised standards. Both the Council (general approach, 13~March~2026) and the Parliament (plenary, 26~March~2026, 569--45--23) rejected the conditional mechanism in favour of fixed dates: 2~December~2027 for stand-alone Annex~III systems and 2~August~2028 for AI embedded in regulated products. The second trilogue on 28~April~2026 ended without agreement. If the trilogue does not close before August~2026, the original deadline applies. It would not, however, close the deficit. The Annex~III, point~2 carve-out is structural: it narrows Article~86 and Article~27 regardless of when the remaining high-risk obligations take effect. The governance gap this paper identifies would persist through any timeline extension, because the gap arises from the Act's architecture, not its implementation schedule.

The Omnibus also proposes a single incident-reporting point intended to reduce duplicative notification under the AI Act, GDPR, and NIS2. For the multi-clock tension addressed by AgentGov-SC's T4 resolution rule, this is a partial response. A unified reporting channel does not resolve the underlying problem that different regimes impose different evidence standards, different scope definitions, and different attribution requirements. Aligning reporting timelines is necessary. Aligning substantive obligations across regimes when an incident spans multiple autonomous systems remains the harder problem.

Institutional guidance clusters around enterprise and general-purpose contexts. The gap is clearest at the municipal level. Singapore's January~2026 governance framework for agentic AI provides a policy benchmark for organisational practice~\citep{imda_2026_agentic_mgf}. System-of-systems settings fall outside its scope. Operational obligations are being clarified ahead of the 2~August~2026 full-applicability date through a General-Purpose AI Code of Practice~\citep{ec_2025_gpai_code} and guidelines on the AI system definition~\citep{ec_2024_ai_act_timeline,ec_2025_ai_system_definition}. Agent security is the subject of a separate NIST Request for Information issued in January~2026~\citep{nist_2026_caisi_rfi,nist_2026_federal_register_rfi}. A governance architecture that cities can activate under real-time agentic autonomy and cross-agency accountability conditions remains absent from all of them.

\subsection{Agentic AI Governance and Smart-City Scholarship}
\label{sec:bg-literature}
 
The governance challenge changes character when the governed object is an agent. Agents pursue goals through sequences of actions, invoking tools and delegating subtasks to other agents as they go~\citep{chan_2024_visibility,kolt_2025_governing_agents}. Risk categories built on static classification struggle with this behaviour. Dimensional governance proposals respond by tracking properties such as decision authority and process autonomy continuously~\citep{engin_hand_2025_dimensional}. Oversight structures face a related mismatch. Episodic approvals and siloed compliance units do not match the operational tempo of systems that adapt between review cycles~\citep{schmitz_2025_oversight_public_sector}.
 
A targeted literature review supports the positioning claim in this paper. The review covered across ACM Digital Library, IEEE Xplore, Scopus, arXiv, and Google Scholar using combinations of ``agentic AI governance,'' ``smart city AI,'' ``multi-agent governance,'' ``autonomous systems governance,'' and ``urban AI accountability'' (2019--2026). The review is not a systematic review or scoping review: inclusion decisions were guided by relevance to the smart-city agentic governance intersection. Exhaustive screening protocols were not applied. Given the 2024--2026 emergence of agentic AI governance as a distinct field, several references are drawn from preprints by leading institutions and are treated as non-peer-reviewed sources with corresponding evidentiary weight. The search identified runtime governance proposals~\citep{wang_2025_mi9}, compliance-mapping frameworks~\citep{eisenberg_2025_ucf}, and municipal AI guidance~\citep{imda_2026_agentic_mgf,helsinki_ai_register}, but no prior work was found that simultaneously integrates smart-city system-of-systems governance, tri-framework compliance mapping, runtime enforcement, resident-facing accountability, and autonomy-calibrated activation for agentic AI. The 11 dimensions in Table~\ref{tab:comparison} are selected to highlight the governance requirements specific to this integration challenge; some prior works cover adjacent capabilities in more operational depth than AgentGov-SC does, particularly MI9 on runtime enforcement and UCF on compliance harmonisation. The comparative gap is therefore partly an artefact of dimension selection. AgentGov-SC's contribution lies in the particular integration and smart-city adaptation of these strands, not in categorical novelty across each dimension.
 
MI9 sets the current benchmark for runtime governance of agentic systems~\citep{wang_2025_mi9}. Its Agency-Risk Index (ARI) computes a $[0,1]$ score from autonomy and adaptability dimensions, with continuity as a separate weighting factor, and maps agents to four governance tiers. Enforcement relies on finite-state conformance checks and continuous activity monitoring, with drift detection triggering escalation when behaviour deviates from approved parameters. What MI9 does not attempt is regulatory traceability. Smart-city critical infrastructure and resident-facing accountability both fall outside its design scope~\citep{wang_2025_mi9}.
 
Cross-framework harmonisation is the focus of the Unified Control Framework (UCF)~\citep{eisenberg_2025_ucf}. Its control set maps obligations under the EU AI Act to corresponding provisions in ISO/IEC~42001 and the NIST AI RMF. The result demonstrates that compliance mapping can be engineered rather than assembled from parallel checklists. Agentic systems fall outside UCF's scope, however. Runtime enforcement primitives and resolution protocols for conflicting obligations under operational pressure are not addressed~\citep{eisenberg_2025_ucf}.
 
Several other works address pieces of the governance problem without closing the intersection targeted here. Enterprise compliance receives attention through Governance-as-a-Service (GaaS), which frames enforcement as a control-plane problem for multi-agent deployments~\citep{gaurav_2025_gaas}. Adversarial resilience is the focus of SAGA's simulation-oriented security architecture~\citep{syros_2025_saga}. Foundational risk-assessment practices~\citep{shavit_2023_practices} and policy-engine approaches to runtime enforcement~\citep{pandey_2025_ssrn_agentic_framework,jackson_2025_ssrn_policy_engine} round out the emerging field. Concurrent work by \citet{nannini_2026_agents_eu_law} provides the most systematic regulatory mapping for AI agent providers under EU law published to date, integrating the AI Act, GDPR, Cyber Resilience Act, NIS2, and Product Liability Directive across nine agent deployment categories. That contribution operates at the provider level across all sectors. The present paper is complementary: it focuses on the deployer-facing accountability gap created by Annex~III, point~2 and specialises the compliance architecture to system-of-systems coordination in smart-city critical infrastructure. Table~\ref{tab:comparison} maps these contributions against 11 governance dimensions. The pattern is consistent: fragmentation across governance strands, not absence of any single strand.

\begin{table*}[htbp]
\centering
\caption{Comparative positioning of AgentGov-SC against key governance frameworks across 11 dimensions. Dimensions reflect the specific governance gap this paper identifies; other frameworks optimise for different objectives (MI9 for runtime enforcement depth, UCF for compliance breadth across regimes). Row~11 (Illustrative Governance Trace) reflects the non-empirical governance trace in Section~\ref{sec:scenario} and the contrasting trace in Section~\ref{sec:scenario-contrast}; no implementation evaluation has been conducted. \checkmark~=~fully addresses; $\circ$~=~partially addresses; --~=~does not address.}
\label{tab:comparison}
\small
\setlength{\tabcolsep}{6pt}
\renewcommand{\arraystretch}{1.2}
\begin{tabular}{@{}p{5cm} c c c c c c c@{}}
\toprule
\textbf{Dimension} & \textbf{MI9} & \textbf{UCF} & \textbf{GaaS} & \textbf{Shavit} & \textbf{SAGA} & \textbf{IMDA} & \textbf{Ours} \\
\midrule
1. Smart City Context        & -- & -- & -- & -- & -- & -- & \checkmark \\
2. Multi-Agent Governance    & $\circ$ & -- & \checkmark & $\circ$ & $\circ$ & $\circ$ & \checkmark \\
3. EU AI Act Mapping         & -- & \checkmark & -- & -- & -- & -- & \checkmark \\
4. ISO 42001 Mapping         & -- & \checkmark & -- & -- & -- & -- & \checkmark \\
5. NIST AI RMF Mapping       & -- & \checkmark & -- & -- & -- & -- & \checkmark \\
6. Runtime Enforcement       & \checkmark & -- & \checkmark & -- & $\circ$ & -- & $\circ$ \\
7. Citizen Accountability    & -- & -- & -- & -- & -- & -- & \checkmark \\
8. Societal/Ethical Analysis & -- & -- & -- & $\circ$ & -- & $\circ$ & \checkmark \\
9. Autonomy-Calibrated       & \checkmark & -- & -- & -- & -- & -- & \checkmark \\
10. Conflict Resolution      & -- & -- & -- & -- & -- & -- & \checkmark \\
\midrule
\textit{11. Illustrative Governance Trace} & \textit{$\circ$} & \textit{--} & \textit{$\circ$} & \textit{--} & \textit{$\circ$} & \textit{--} & \textit{$\circ$} \\
\bottomrule
\end{tabular}
\end{table*}

Urban AI is a governance problem before it is a technical one. Kitchin's account of the real-time city describes how pervasive sensing shifts the locus of power toward those who control data pipelines~\citep{kitchin_2014_realtime,kitchin_2016_ethics}. Transparency in municipal algorithmic systems is constrained by procurement rules and vendor secrecy. Internal capacity to audit is limited~\citep{brauneis_goodman_2018_transparency}. What counts as a problem depends on control-room practices and on whose knowledge those practices make visible~\citep{coletta_kitchin_2017_algorhythmic}. AI complicates the picture. Discretion now flows across agencies and contractors through automated pipelines, and existing accountability frameworks were not designed to follow it~\citep{goldsmith_yang_2024_accountability,cardullo_kitchin_2019_citizenship}.

Guideline initiatives converge on values. Enforcement specifics remain absent~\citep{jobin_2019_global,hagendorff_2020_ethics_of_ai_ethics,mittelstadt_2019_principles}. Abstraction traps explain one mechanism. Fairness and accountability concepts detach from the social setting that gives them meaning, then break at system boundaries~\citep{selbst_2019_abstraction}. The structural power asymmetries identified in Section~\ref{sec:introduction} are amplified in this context~\citep{crawford_2021_atlas,benjamin_2019_race}. Frameworks that centre community knowledge in governance design become correspondingly more important~\citep{birhane_2021_algorithmic_injustice}. AgentGov-SC responds by treating resident-facing accountability as an operational requirement implemented through City Layer mechanisms.
 
A missing bridge is now visible. Much of the smart-city governance literature matured around predictive analytics and dashboard-driven oversight, while agentic systems introduce autonomy, tool use, and multi-step action as the default interaction pattern. Recent smart-city work has begun to incorporate multi-agent systems for urban decision support, yet it evaluates performance, not governance controls~\citep{kalyuzhnaya_2025_llm_agents}. Agentic governance proposals rarely engage the distinctive accountability conditions of cities~\citep{sawhney_2023_contestations}. AgentGov-SC is positioned as a response to that gap: a governance architecture that treats the city context as the binding constraint on agentic autonomy. The city is not an incidental deployment setting.

\section{The AgentGov-SC Framework}
\label{sec:framework}
 
The regulatory and governance gaps identified in Section~\ref{sec:background} define the design space for AgentGov-SC: the EU AI Act's bounded-system assumption, the Annex~III, point~2 citizen-accountability carve-out, and the absence of any existing framework addressing smart-city agentic governance with concurrent compliance mapping. The framework separates governance into three layers and links governance intensity to societal impact. Conflict-resolution rules for cross-framework tensions complete the design. Twenty-five governance measures (R-01 through R-25) operationalise this design. Each is mapped to regulatory obligations and assigned to a governance layer. The full control catalog appears in Table~\ref{tab:controls} (Section~\ref{sec:compliance}), structured for bidirectional traceability. The subsections below introduce measures by function, not by number, and each identifier is accompanied by a descriptive name at first mention.

\subsection{Three-Layer Architecture}
\label{sec:architecture}

\begin{figure*}[!t]
\centering
\includegraphics[width=\textwidth]{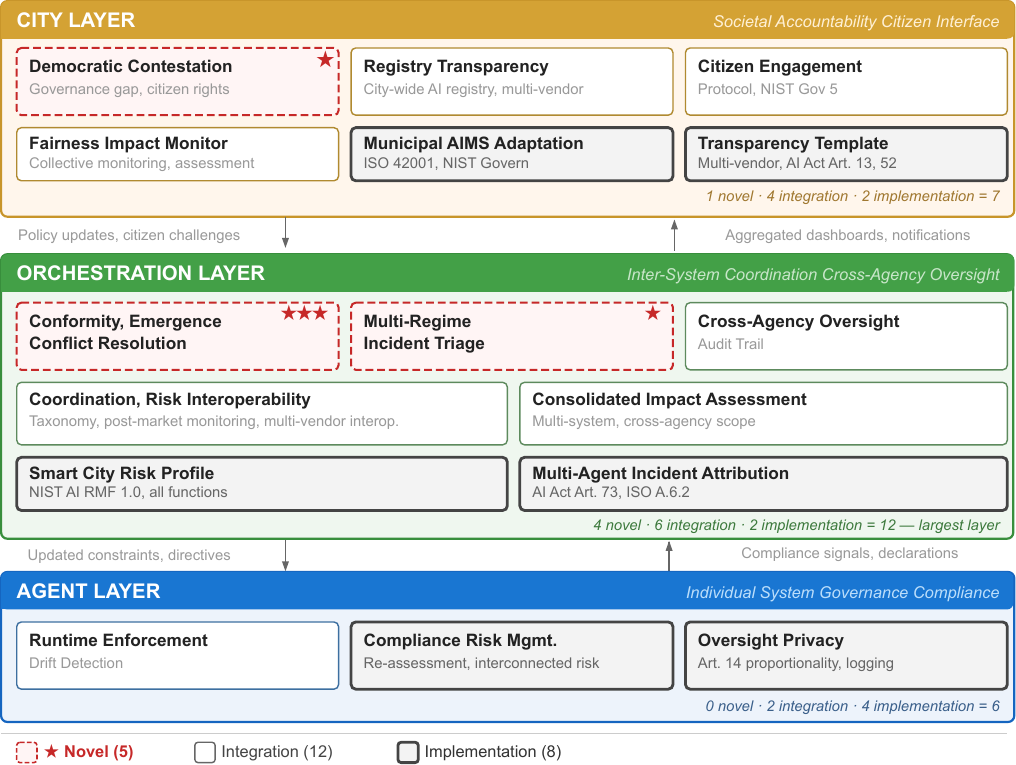}
\caption{AgentGov-SC three-layer governance architecture. Layers separate unit-level compliance (Agent), inter-system coordination (Orchestration), and societal accountability (City). Regulatory framework traces shown on margins; novel contributions marked with asterisk~(*).}
\label{fig:architecture}
\end{figure*}

AgentGov-SC separates governance into three layers because failures in smart-city deployments rarely stay contained to a single model. A resident affected by a corridor-level decision may never know which of three vendors' systems contributed to it, because sensors and data pipelines cross organisational boundaries that accountability structures were not built to follow~\citep{kitchin_2016_ethics,coletta_kitchin_2017_algorhythmic,brauneis_goodman_2018_transparency}. The architecture therefore tracks the scale at which a governance intervention can actually be executed and evidenced. At the lowest level, the Agent Layer governs an individual AI system as an operational unit. Above it, the Orchestration Layer addresses the space between systems, where coordination failures and propagation risks originate. The City Layer sits at the top, facing outward. Cumulative effects become visible here through contestation, transparency, and public oversight. Figure~\ref{fig:architecture} summarises the layers and their escalation pathways. A fourth layer is not introduced because the remaining governance needs identified in this paper collapse into one of these execution contexts: unit-level controls, interaction controls, or societal accountability controls.

\paragraph{Agent Layer.} The Agent Layer governs individual AI systems that must produce governance-ready evidence at runtime, before a city-level process can detect drift or impact. AgentGov-SC groups unit-level mechanisms around three functions that tend to be conflated in practice. Behavioural boundary enforcement comes first. Runtime policy enforcement and drift detection (R-09, R-10) establish a monitored envelope for action, with triggers that turn silent degradation into an explicit governance event. Assurance validity is a separate concern. Interconnected risk management and conformity re-assessment (R-20, R-23) treat operational change, integration change, and environment change as triggers for reassessment, not as background noise. Defensible evidence generation under privacy constraints rounds out the Agent Layer. Privacy-preserving logging (R-19) ensures that auditability does not default to maximal collection. Retention is bounded and access is tiered. Oversight functions can reconstruct decision pathways without requiring full personal-data exposure. MI9's conformance checking and monitoring techniques align naturally with this layer. The Agent Layer binds runtime signals to explicit traceability and requires a standardised declaration-and-alert interface so that unit-level evidence can be consumed by the Orchestration Layer when interaction governance becomes the controlling concern~\citep{wang_2025_mi9}.

\paragraph{Orchestration Layer.} The Orchestration Layer governs interaction effects across autonomous systems that share infrastructure, data, and operational authority, the coupling that unit-level assurance cannot see. AgentGov-SC makes this inter-system boundary explicit and assigns it primary governance responsibility. Two requirements carry most of the conceptual load. R-01 introduces system-of-systems conformity assessment as an obligation for smart-city autonomy. Its purpose is to define and continuously validate the permitted interaction topology: which systems can exchange signals with which others, what data flows are permissible, and what joint behaviours are disallowed. Safety-critical dependencies receive separate treatment. The concept draws on established systems-of-systems safety engineering, particularly Leveson's STAMP framework, which reframes emergent risk in coupled systems as a control-structure problem. Component failure is not the primary concern; the coupling is~\citep{leveson_2011_engineering_safer_world}. R-01 applies that insight to AI governance by making the interaction topology an assessable and auditable regulatory object. Collective emergent impact assessment (R-03) complements this by treating propagation as an assessable risk object, focusing on second-order effects that arise only when systems co-adapt or co-act over shared resources. The Orchestration Layer carries the largest share of governance duties in the requirement mapping (12 of 25), reflecting the authors' analytical judgment that the interaction space is the principal governance bottleneck in smart-city infrastructure. Supporting mechanisms make this layer executable. Cross-agency human oversight coordination (R-06) operates under shared escalation triggers, while an inter-agent communication audit trail (R-07) retains a reconstructible record across vendors. When multiple agents are involved in an incident, attribution (R-24) assigns responsibility across the vendors and authorities concerned.

\paragraph{City Layer.} The City Layer addresses a different failure mode: resident-facing consequences can be pervasive even when internal compliance artefacts are technically correct~\citep{cardullo_kitchin_2019_citizenship}. AgentGov-SC treats this as a governance requirement. A democratic contestation mechanism (R-04) is introduced as a city-scale control that operationalises contestation concepts from the smart-city governance literature~\citep{sawhney_2023_contestations} as an auditable requirement with regulatory traceability. It provides a structured pathway for residents to challenge outcomes that plausibly arise from cross-system interaction without forcing those challenges into a single-agency silo. The mechanism is procedural and evidence-oriented: it requires that affected individuals can obtain a usable account of the decision taken and the information materially relied upon. It must also specify what review is available and what remedy is possible. The remainder of the City Layer is enabling infrastructure for accountability. Visibility comes through a city-wide AI registry and multi-vendor transparency aggregation (R-13, R-14). Participatory engagement procedures (R-16) keep the registry usable in repeated interactions. Collective fairness monitoring and consolidated impact assessment (R-08, R-12) close the loop by tracking cumulative effects of incidents and interactions across systems. A resident whose fine resulted from a cascade across three systems needs an assessment that reflects all three. Single-system performance reviews cannot provide that. Power and accountability concerns raised by smart-city critiques~\citep{brauneis_goodman_2018_transparency} are addressed here through implementable controls.

Urban studies scholarship has named the automation-to-autonomy transition directly. Cugurullo's account of the smart city argues that AI is producing urban systems whose decisions humans neither guide nor supervise~\citep{cugurullo_2020_urban_ai,cugurullo_2021_frankenstein}. Subsequent work extends the argument: AI urbanism is not the technological successor of smart urbanism but a different urban condition in which agency is distributed across human and non-human actors~\citep{cugurullo_etal_2024_ai_urbanism}. A recent typology by \citet{tiwari_2025_agentic} distinguishes automation, autonomy, and agency as analytically distinct properties of urban AI systems. AgentGov-SC's autonomy-calibrated activation model (Section~\ref{sec:autonomy}) treats that progression as a governance activation rule: at each level, the model specifies which governance mechanisms must be in scope and what evidence must be produced.

\noindent
Inter-layer protocols connect the layers through escalation and feedback, not symmetric mirroring. A typical escalation starts with a unit-level trigger: a drift threshold breach (R-10) or a policy violation signal (R-09) that only becomes meaningful in context. The signal moves upward carrying a declaration of dependencies and current operating mode. The Orchestration Layer checks whether interaction constraints still hold (R-01) and whether propagation risks are building (R-03). Where agencies need to coordinate, R-06 sets the joint oversight path and R-07 keeps the interaction record intact for any future incident reconstruction. When consequences reach residents, the City Layer takes over: what gets disclosed and who gets notified. The contestation route (R-04/R-13/R-16) opens at this level. Downward flow compresses into enforceable updates. Contestation outcomes and policy revisions flow downward as orchestration directives. Revised interaction constraints become system-level configuration changes.

\subsection{Autonomy-Calibrated Governance Model}
\label{sec:autonomy}

Agentic systems vary not only in technical capability but also in the kinds of societal consequences their autonomy can produce. MI9's Agency-Risk Index (ARI) provides a technically grounded lens for runtime governance by mapping agent capability to containment-oriented governance tiers~\citep{wang_2025_mi9}. AgentGov-SC is designed to sit alongside ARI, not to replace it. The autonomy-calibrated governance model asks how much governance is warranted given what an agent does to residents and urban infrastructure, using impact-facing dimensions. Population affected and decision reversibility anchor the assessment. Infrastructure criticality, rights implications, and the likelihood of cross-system propagation determine whether the Orchestration Layer must activate. The approach aligns with the EU AI Act's risk-management framing and its proportionality requirement for human oversight in Article~14, which links oversight design to the risks posed by the high-risk system and the circumstances of its use~\citep{euaiact_2024}. It also reflects \citet{shneiderman_2022_hcai}'s argument that high levels of automation and high levels of human control are not mutually exclusive, and that governance structures should be calibrated to the operational context rather than imposed as uniform constraints.

Table~\ref{tab:govlevels} summarises the autonomy-calibrated governance levels as an activation rule. Each level links impact-facing conditions to the minimum set of AgentGov-SC layers that must be in scope and the corresponding human-oversight posture required under Article~14 proportionality. The UAE examples are illustrative anchors drawn from publicly documented deployments, G5 represents a plausible city-scale trajectory, not a documented current deployment.

\begin{table*}[htbp]
\centering
\caption{Autonomy-calibrated governance levels as an activation rule. Levels escalate governance intensity based on societal impact and specify which AgentGov-SC layers that must be in scope. Oversight postures are aligned to Article~14 proportionality. UAE references are illustrative anchors; G5 is forward-looking. Layer codes: A~=~Agent; O~=~Orchestration; C~=~City. Activation: Full~=~fully active; Basic~=~basic activation; Off~=~inactive.}
\label{tab:govlevels}
\footnotesize
\setlength{\tabcolsep}{4pt}
\renewcommand{\arraystretch}{1.3}
\begin{tabular}{@{}c L{3.5cm} L{4cm} L{2.8cm} L{3cm}@{}}
\toprule
\textbf{Level} & \textbf{Scope} & \textbf{Art.~14 Oversight Posture} & \textbf{Layers Active} & \textbf{UAE Reference} \\
\midrule
\rowcolor{gray!6}
\makecell[c]{\textbf{G1}\\{\scriptsize Advisory}} &
AI provides recommendations; human authorises execution. &
Human-in-command; routine review with clear responsibility. &
A: Full; O: Off; C: Off &
Falcon Eye (Surveillance) \\[4pt]
\makecell[c]{\textbf{G2}\\{\scriptsize Assistive}} &
AI executes bounded tasks under human review within a defined workflow. &
Human-on-the-loop; approval gates and auditable task boundaries. &
A: Full; O: Off; C: Off &
Rammas (utility customer support) \\[4pt]
\rowcolor{gray!6}
\makecell[c]{\textbf{G3}\\{\scriptsize Semi-autonomous}} &
AI makes operational decisions in a defined scope; humans intervene via escalation. &
Human-over-the-loop; intervention protocols and coordinated escalation. &
A: Full; O: Basic; C: Off &
Oyoon (citywide sensing and enforcement) \\[4pt]
\makecell[c]{\textbf{G4}\\{\scriptsize Autonomous-critical}} &
AI operates autonomously in critical infrastructure; supervisory override. &
Supervisory oversight proportionate to criticality; emergency override and cross-agency coordination. &
A: Full; O: Full; C: Basic &
DEWA GTIC (gas turbine intelligent control) \\[4pt]
\rowcolor{gray!6}
\makecell[c]{\textbf{G5}\\{\scriptsize Autonomous-systemic}} &
Multi-agent ecosystem spans agencies and vendors; emergent behaviour from city-wide interaction. &
Societal oversight and contestability; citywide accountability as part of risk treatment. &
A: Full; O: Full; C: Full &
Forward-looking city-scale ecosystem (not currently documented) \\
\bottomrule
\end{tabular}
\end{table*}

The activation rule matters more than the taxonomy. The model determines which governance mechanisms must be in scope for a deployment, and it makes that decision defensible through an explicit proportionality rationale. The boundary between G3 and G4 is a practical pivot. Once autonomy is exercised inside critical infrastructure, reversibility narrows and the cost of coordination failure rises, because incidents propagate into safety, continuity of essential services, and resident trust. The contrast between Oyoon's operational influence in public space and DEWA's autonomous control in energy infrastructure illustrates why governance intensity cannot be derived from capability alone. Some agents can be technically sophisticated yet remain advisory in effect, while a comparatively simple controller can demand higher governance because its failure modes are infrastructural rather than transactional. MI9's ARI and the autonomy-calibrated model therefore answer different questions. ARI helps estimate technical agency and containment needs~\citep{wang_2025_mi9}. The autonomy-calibrated model answers a different question: how much governance burden does the societal impact of this deployment warrant? Organisational management and risk-management obligations under ISO/IEC~42001 and the NIST AI RMF apply at every level~\citep{isoiec_42001_2023,nist_2023_airmf}. What changes is the evidence burden. At G2, a single agency can produce the required artefacts. At G4, coordination across agencies becomes the binding constraint. Level assignment is qualitative in this version of the framework; the model functions as a structured decision aid requiring professional judgment, not as a reproducible scoring instrument. To reduce subjectivity at the most consequential boundary, the G3/G4 threshold can be operationalised through a two-part test: (i)~does the system exercise autonomous control within infrastructure whose failure directly endangers persons or essential services, and (ii)~does the system's operational scope create dependencies that cross organisational boundaries? Where both conditions are met, G4 governance applies and the Orchestration Layer activates in full. Disputes about level assignment should be resolved through the consolidated impact assessment process (R-12), which provides the evidentiary basis for proportionality determinations. Validating a quantitative scoring instrument for consistent classification across cities is left to future work.

\subsection{Tri-Framework Compliance Mapping}
\label{sec:compliance}

AgentGov-SC treats compliance as a design problem that must survive runtime autonomy and inter-system coordination, not a documentation exercise performed after deployment. The mapping is designed to support bidirectional traceability. An engineer starts from a regulatory obligation, translates it into a governance measure (R-xx), and operationalises it as one or more controls (AG-xxx) assigned to a layer. An auditor enters from the opposite end, tracing any control back to the obligation it satisfies. Full reverse traceability through evidence artefacts to specific legal coverage would require operational demonstration that this version does not yet provide; the mapping should be read as structured for bidirectional use, not verified in both directions. The tri-framework mapping~\citep{euaiact_2024,isoiec_42001_2023,nist_2023_airmf} builds on UCF's precedent that cross-regime harmonisation can take the form of a unified control set~\citep{eisenberg_2025_ucf}, specialised here to agentic autonomy in smart-city critical infrastructure.

Table~\ref{tab:controls} summarises the five novel governance measures that respond directly to the accountability gaps identified in Sections~\ref{sec:introduction} and~\ref{sec:background}. Each measure addresses a governance need that no existing framework covers: system-of-systems conformity, cross-framework conflict resolution, collective emergent impact assessment, democratic contestation, and multi-regime incident triage. The full 25-measure control catalog, including 12 integration measures and 8 implementation measures with bidirectional regulatory traceability, is provided in Supplementary Material~S1.

\begin{table*}[!t]
\centering
\caption{Novel governance measures introduced by AgentGov-SC. Each addresses a governance gap not covered by the EU AI Act, ISO/IEC~42001, or the NIST AI RMF. The full 25-measure control catalog with bidirectional regulatory traceability is in Supplementary Material~S1.}
\label{tab:controls}
\footnotesize
\setlength{\tabcolsep}{4pt}
\renewcommand{\arraystretch}{1.4}
\begin{tabular}{@{} l L{3.2cm} L{4.5cm} L{4.5cm} l @{}}
\toprule
\textbf{ID} & \textbf{Governance Measure} & \textbf{Accountability Gap Addressed} & \textbf{Legal Grounding} & \textbf{Layer} \\
\midrule

R-01 & System-of-systems conformity assessment & No existing framework assesses the interaction topology between autonomous systems operated by separate authorities & Extends Art.~43 conformity assessment to interaction-level risk; draws on STAMP~\citep{leveson_2011_engineering_safer_world} & Orchestration \\[4pt]
R-02 & Cross-framework conflict resolution & Competing obligations across EU AI Act, ISO/IEC~42001, and NIST AI RMF are not reconciled at runtime & Operationalises five tension-specific rules (Table~\ref{tab:conflicts}) with evidence generation & Orchestration \\[4pt]
\rowcolor{gray!4}
R-03 & Collective emergent impact assessment & Art.~9 risk management scopes to individual systems; no provision for propagation effects across coupled agents & Extends Art.~9(2)(b) to second-order effects that arise only when systems co-adapt over shared resources & Orchestration \\[4pt]
R-04 & Democratic contestation mechanism & Art.~86 explanation rights exclude Annex~III, point~2 systems; no cross-agency contestation pathway exists & Operationalises contestation concepts~\citep{sawhney_2023_contestations} as an auditable, cross-boundary right & City \\[4pt]
\rowcolor{gray!4}
R-05 & Multi-regime incident triage & NIS2 (24h), GDPR (72h), and AI Act notification windows conflict; no unified triage protocol for cross-agency incidents & Aligns to strictest-clock baseline with parallel notifications and shared incident record & Orchestration \\
\bottomrule
\end{tabular}
\end{table*}

The mapping supports multiple entry points. A procurement officer traces Article~14 proportionality to the relevant measure and verifies that a vendor's proposal specifies what proportionate oversight means. A resident challenging an enforcement decision traces which governance measures were or were not satisfied and directs contestation to the responsible layer.

The remaining 20 measures fall into two categories. Twelve integration measures combine provisions from multiple frameworks into operational controls (cross-agency oversight, inter-agent audit trails, consolidated impact assessment, city-wide AI registry, and others). Eight implementation measures operationalise existing obligations for the smart-city context (privacy-preserving logging, interconnected risk management, conformity re-assessment). All 25 are specified with bidirectional regulatory traceability in Supplementary Material~S1.

The five novel measures vary in their character. R-01 and R-03 introduce new governance objects (interaction topologies and propagation effects) drawing on systems-of-systems safety antecedents~\citep{leveson_2011_engineering_safer_world}. R-02, R-04, and R-05 operationalise known concepts (cross-framework tension management~\citep{eisenberg_2025_ucf}, urban contestation~\citep{sawhney_2023_contestations}, multi-regime coordination~\citep{nis2_2022_2555}) by giving them procedural specification, regulatory traceability, and layer assignment.

\subsection{Cross-Framework Conflict Resolution}
\label{sec:conflicts}

Cross-framework tensions in smart-city deployments are rarely legal contradictions in a strict sense. They turn operational when compliance teams face different timelines, engineers face different evidence expectations, and incident responders face disclosure constraints that cut across all three frameworks. Real-time autonomy does not wait for frameworks to align. AgentGov-SC therefore treats conflict resolution as a governance mechanism not as a documentation footnote. The protocol operationalises R-02 (cross-framework conflict resolution) through five tension-specific rules.

Table~\ref{tab:conflicts} translates each tension into a resolution rule. Each rule is assigned to a responsible layer and tied to controls that generate evidence. That evidence must survive audit, incident review, and citizen challenge. UCF usefully surfaces that tensions exist across regimes, but it stops short of specifying how competing obligations should be reconciled during operation~\citep{eisenberg_2025_ucf}.

\begin{table*}[htbp]
\centering
\caption{Cross-framework conflict resolution protocol. Each operational tension identifies conflicting regulatory requirements, specifies AgentGov-SC's resolution mechanism, assigns implementation responsibility to a layer, and grounds resolution through UAE deployment context. Acronyms: FRIA~=~Fundamental Rights Impact Assessment; DPIA~=~Data Protection Impact Assessment; ISMS~=~Information Security Management System.}
\label{tab:conflicts}
\footnotesize
\setlength{\tabcolsep}{4pt}
\renewcommand{\arraystretch}{1.4}
\begin{tabular}{@{}c L{2cm} L{3cm} L{6.5cm} l L{3.5cm}@{}}
\toprule
\textbf{ID} & \textbf{Tension} & \textbf{Frameworks} & \textbf{Resolution Mechanism} & \textbf{Layer} & \textbf{UAE Context} \\
\midrule
\rowcolor{gray!4}
\textbf{T1} &
Logging vs.\ minimization &
AI Act record-keeping vs. GDPR Art.~5(1)(c) &
R-19: Privacy-preserving logging with structured minimisation, tiered access, auditable scope &
Agent &
\textit{Oyoon}: large-scale sensing (300K+ cameras) vs. minimisation pressures \\[3pt]
\textbf{T2} &
Erasure vs.\ retention &
GDPR Art.~17 vs. ISO/IEC 42001 documentation &
R-19/R-07: Graduated retention with time-bounded schedules, pseudonymisation, purpose-limited audit &
Orchestration &
Citizen service records: auditability + deletion pathways \\[3pt]
\rowcolor{gray!4}
\textbf{T3} &
Assessment proliferation &
FRIA, DPIA, ISMS, AIMS, RMF overlaps &
R-12: Consolidated methodology with regime-specific traceability outputs &
Orchestration &
DEWA: unified approach across programs + standards \\[3pt]
\textbf{T4} &
Incident timelines &
NIS2 (24h) / GDPR (72h) / AI Act windows &
R-05: Coordinated triage with strictest-clock baseline, parallel notifications, shared record &
Orchestration &
Multi-agency incidents under unified time pressure \\[3pt]
\rowcolor{gray!4}
\textbf{T5} &
Transparency vs.\ trade secrets &
AI Act transparency vs. TRIPS/IP protection &
R-14: Tiered disclosure (public vs. regulator/auditor) under confidentiality &
City &
Multi-vendor platforms: IP protection + regulatory visibility \\
\bottomrule
\end{tabular}
\end{table*}

Two tensions recur in city deployments because they sit on the critical path of routine operation. Logging versus minimisation (T1) is the more pervasive. Oversight and contestation require a reconstructible account of an incident, whereas data protection regimes emphasise minimisation and purpose limitation~\citep{gdpr_2016_679}. Large-scale sensing programs intensify this trade-off. Where infrastructure-level monitoring is described at the scale of hundreds of thousands of cameras~\citep{gulfnews_2021_oyoon_300k}, an accountability posture that simply ``logs everything'' collapses into ungovernability, while a posture that logs too little makes incident reconstruction and redress performative. The multi-clock incident problem (T4) is the more time-sensitive. Cybersecurity, privacy, and AI governance regimes impose different notification expectations, and a cross-agency incident can stall when responders lack a single operational rule for triage under time pressure~\citep{nis2_2022_2555,gdpr_2016_679}. This is amplified in smart cities because incidents do not respect organisational boundaries. A failure that begins as a technical fault can quickly acquire public-safety and rights implications once enforcement or access decisions are triggered. Other tensions remain important but are less frequently decisive in the first hour of response. Retention versus erasure (T2), assessment proliferation (T3), and transparency versus trade secrets (T5) are handled through the same protocol and are mapped in Table~\ref{tab:conflicts}.

Resolution follows two principles: meet the strictest applicable clock for time-sensitive duties, and partition evidence so accountability does not require maximal disclosure. The logging-minimisation tension (T1) is resolved through R-19, which specifies privacy-preserving logging with structured minimisation and auditable scope controls. Incident timelines (T4) demand a different mechanism: R-05 coordinates triage across regimes by aligning operational response to the strictest reporting timeline while preserving a single incident record. Assessment proliferation (T3) and transparency-versus-trade-secrets (T5) are handled through consolidation and partitioning respectively. R-12 provides one methodological spine that can be traced to multiple regimes without multiplying inconsistent artefacts. R-14 separates public disclosure from regulator and auditor access under confidentiality protections~\citep{trips_1994}. These mechanisms are not presented as universal legal solutions. They are operational rules intended to keep governance executable when autonomy and cross-agency coupling are the default.

\section{Illustrative Application: UAE Smart City Infrastructure}
\label{sec:application}

To test whether AgentGov-SC's categories correspond to real governance conditions, one needs a city where autonomous AI already operates at infrastructure scale across multiple authorities. The UAE provides that setting. Ten publicly documented systems span transport and energy in Dubai and Abu Dhabi, with surveillance and digital government adding further complexity. Autonomy reaches Level~4 in critical energy infrastructure. The scenario that follows is forward-looking: it projects current systems one step beyond their documented autonomy to examine the governance demands that agentic scaling will impose.

One observation frames the entire analysis. The UAE has built institutional coordination structures that most smart city ecosystems lack. A cross-agency supervisory committee seats police, energy, and transport authorities at the same table alongside municipal entities. A binding legal framework governs autonomous vehicles. Emergency coordination protocols span all government entities. Yet none of these mechanisms currently extends to real-time governance of autonomous system interactions. The gap is not institutional absence. It is a scope mismatch between mature coordination bodies and the emerging operational demands of agentic AI.

Autonomy-level assignments in Table~\ref{tab:uae-inventory} follow an evidence protocol based on three criteria derived from public sources. The first is the scope of automated decision-making documented in operator statements, ranging from advisory through bounded-task and operational to real-time control. The second is the nature of human involvement described, whether pre-approval, exception-handling, monitoring, or supervisory override. Domain criticality provides the third criterion: customer-facing, public-space, or critical infrastructure contexts carry different governance weight. Each system was classified by the authors on the basis of the public evidence cited in Section~\ref{sec:uae-systems}. The protocol is qualitative, without inter-rater reliability testing. Adjacent levels could reasonably be assigned differently by independent assessors, particularly at the L2/L3 boundary. There, the distinction between human-assisted and bounded-autonomous operation turns on operational details that public sources do not fully reveal.

\subsection{UAE Smart City AI Systems}
\label{sec:uae-systems}

Ten publicly documented AI systems across Dubai and Abu Dhabi are inventoried in Table~\ref{tab:uae-inventory}. Autonomy levels span L2 (AI assistance) through L4 (fully autonomous real-time control). Inclusion requires evidence of non-trivial automated decision-making in an urban function.

Three systems warrant brief comment. RTA's UTC-UX Fusion manages adaptive signal control across approximately 300~intersections~\citep{rta_2025_utcux,dmo_2025_rta_signal_upgrade}. Dubai Police's Oyoon program aggregates more than 300,000~cameras~\citep{gulfnews_2021_oyoon_300k}. DEWA's GTIC presents the highest documented autonomy: reinforcement learning controllers on nine gas turbines at Jebel Ali, co-developed with Siemens Energy, autonomously optimising combustion~\citep{dewa_2022_gtic_mstation,siemens_energy_gtic_story}. The governance question is what happens when autonomous energy decisions interact with dependent infrastructure that no single authority oversees.

Operational overlap is already visible. RTA and Dubai Police run a joint Traffic Incidents Management Unit across 17 corridors and 951~km~\citep{dmo_2024_rta_police_traffic_unit}. The Oyoon Supervisory Committee brings together Dubai Police, DEWA, RTA, Dubai Municipality and five other government entities~\citep{dubai_execres_34_2021_oyoon}. All government entities are required to cooperate under Article~10 of the Resolution No.~(34) of 2021. Cross-agency coordination takes its formalised shape in the autonomous vehicle domain. RTA's coordination authority for autonomous vehicle operations is grounded in Dubai Law No.~(9) of 2023~\citep{dubai_law_9_2023_av,sheikhmohammed_2023_av_law}.

\begin{figure*}[!t]
\centering
\includegraphics[width=\textwidth]{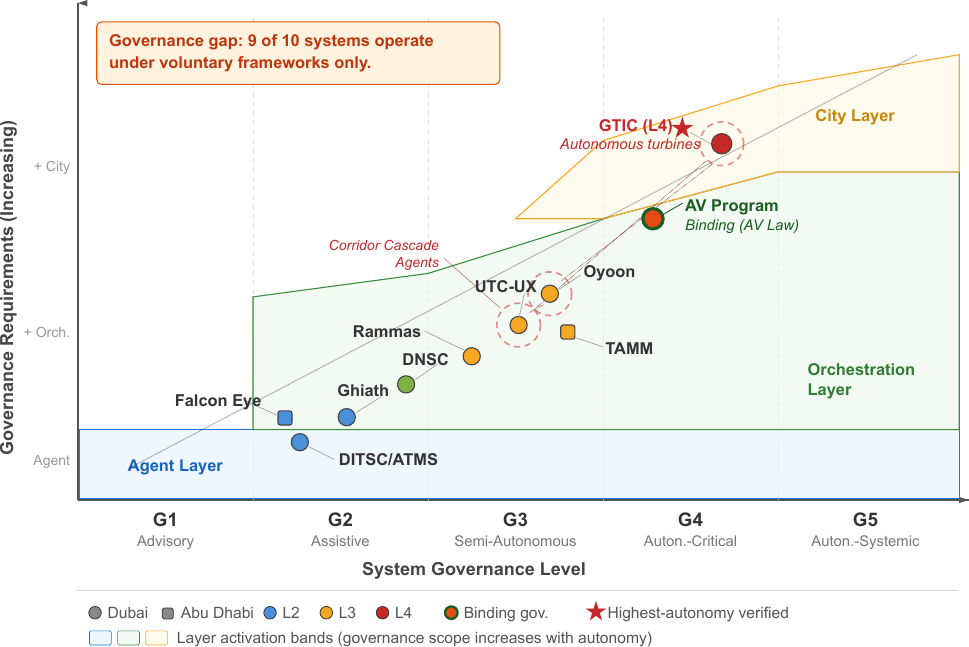}
\caption{Autonomy-calibrated governance model plotting publicly documented UAE smart city AI systems against measures. Unlike MI9's Agency-Risk Index (ARI), which measures technical agent capability, AgentGov-SC calibrates governance measures based on societal impact. Shaded regions indicate which layer mechanisms activate at each governance level. DEWA GTIC (G4) represents the highest-autonomy publicly documented deployment.}
\label{fig:autonomy}
\end{figure*}

Figure~\ref{fig:autonomy} maps all ten systems against AgentGov-SC governance levels. Layer activation scales with societal impact. Nine of the ten systems operate under voluntary frameworks. The only exception is the AV program. This ratio is not a UAE-specific finding. It is the default condition wherever cities deploy autonomous systems faster than governance institutions can keep pace.

\begin{table*}[htbp]
\centering
\caption{UAE smart city AI systems identified from public sources. Autonomy levels are author-assigned per the Evidence Protocol (Section~\ref{sec:application}). Only the AV Program operates under binding AI-specific governance. Autonomy levels: L2~=~human decides, AI assists; L3~=~bounded autonomy, human handles exceptions; L4~=~autonomous real-time control, human monitors. Governance: Binding~=~binding AI-specific governance; Voluntary~=~voluntary or no AI-specific governance. All key metrics are drawn from public operator statements and reporting cited in Section~\ref{sec:uae-systems}.}
\label{tab:uae-inventory}
\footnotesize
\setlength{\tabcolsep}{5pt}
\renewcommand{\arraystretch}{1.25}
\begin{tabular}{@{} L{2.2cm} L{3cm} l l l L{5cm} @{}}
\toprule
\textbf{System} & \textbf{Authority} & \textbf{Domain} & \textbf{Autonomy} & \textbf{Governance} & \textbf{Key Metric} \\
\midrule
UTC-UX Fusion      & RTA              & Traffic         & L3        & Voluntary  & 300 intersections \\
\rowcolor{gray!4}
DITSC/ATMS         & RTA              & Traffic         & L2        & Voluntary  & AI-assisted incident management, AI advisory \\
AV Program         & RTA              & Transport       & L3--L4    & Binding    & Testing phase (SAE L4) \\
\rowcolor{gray!4}
Oyoon              & Dubai Police     & Surveillance    & L3        & Voluntary  & 300,000+ cameras \\
Ghiath             & Dubai Police     & Patrol          & L2        & Voluntary  & AI-assisted patrols \\
\rowcolor{gray!4}
GTIC               & DEWA             & Energy          & L4        & Voluntary  & Autonomous turbines \\
DNSC               & DEWA             & Energy Distrib. & L2--L3    & Voluntary  & Smart grid monitoring \\
\rowcolor{gray!4}
Rammas             & DEWA             & Customer        & L3        & Voluntary  & 12.7M+ inquiries \\
TAMM/AutoGov       & Abu Dhabi DGE    & Gov.\ Services  & L3        & Voluntary  & 3.6M users \\
\rowcolor{gray!4}
Falcon Eye         & Abu Dhabi ADMCC  & Surveillance    & L2        & Voluntary  & 45,000+ sensors \\
\bottomrule
\end{tabular}
\end{table*}

\subsection{UAE AI Governance Environment}
\label{sec:uae-governance}

The UAE's national AI Charter~\citep{uae_ai_charter_2024} sets responsible-AI objectives at the federal level. The objectives are converted into practical guidance for deploying organisations by Dubai's Ethical AI Toolkit. It does not impose legally enforceable duties~\citep{digitaldubai_2019_ai_ethics_toolkit}. DEWA has publicly adopted it across its AI projects~\citep{dewa_2020_ethical_ai_toolkit}. The instruments are not decorative. Common language across stakeholders and shared procurement norms are governance outcomes that binding regulation struggles to produce on its own.

Coordination between agencies goes further. Oyoon Supervisory Committee's mandate covers AI security policy and surveillance system integration, and all government entities are required to cooperate. Its mandate covered standards and policy, not operational deconfliction between autonomous systems. Cross-infrastructure coordination authority is addressed for emergency management~\citep{dubai_decree_4_2021_sccdm}. These bodies provide institutional architecture that most cities lack. The challenge is not capacity. It is extending coordination mandates from policy oversight and emergency response to the operational tempo of autonomous system interactions.

Personal data law illustrates where the governance boundary currently falls. Federal Decree-Law No.~(45) of 2021 excludes government data and governmental entities from its scope under Article~2(2)~\citep{uae_pdpl_2021}. Surveillance and transport enforcement both involve government entities processing resident data at scale. Municipal services add a third category. All three fall outside the reach of the primary data protection framework.

A different model operates within the same jurisdiction. DIFC Regulation~10 requires entities deploying autonomous and semi-autonomous processing systems to implement safeguards proportionate to the risk of harm, including human review and mechanisms for affected individuals to challenge outcomes~\citep{difc_regulation10_2023,difc_dp_regs_2023}. AgentGov-SC's contestation mechanism (R-04) generalises this precedent from a financial-zone perimeter to the cross-agency urban fabric. The UAE's systems are operational and its coordination bodies exist. Extending coordination mandates to the operational tempo of autonomous infrastructure is the remaining step.

\subsection{Illustrative Governance Trace: The Corridor Cascade}
\label{sec:scenario}
 
A substation fire on 20~December 2025 knocked out power across several San Francisco districts. Traffic signals went dark. Waymo's autonomous robotaxis stalled in intersections blocking arterial routes~\citep{techcrunch_2025_waymo_sf_blackout}. The event illustrates cross-domain cascade vulnerability, but not the agentic feedback dynamics that the scenario below adds. In San Francisco, the autonomous vehicles did not adaptively re-optimise; they stalled. The governance challenge examined here arises when systems actively adapt to each other's actions. The coupling paths that result emerge at runtime. Fixed infrastructure dependencies are visible at design time. Agentic coupling paths are not.
 
The scenario that follows is set in a near-future deployment window where three documented systems have each advanced one step beyond their current operational autonomy. UTC-UX Fusion has completed rollout to its full 300-intersection target with network-level adaptive optimisation. DEWA's demand-side management has progressed from tariff-based conservation to automated distribution-level demand response, consistent with the trajectory set out in the Dubai Supreme Council of Energy's DSM Strategy 2050~\citep{dubaisce_2024_dsm_strategy}. Oyoon's enforcement pipeline has moved toward faster automated processing with reduced human verification latency. Each projection is modest and consistent with publicly stated trajectories. The scenario is explicitly hypothetical. No claim is made that these interactions have occurred or that current operational practice is inadequate. The section is an illustrative governance trace, not an empirical validation or operational evaluation.

\paragraph{Setting and agents.}
The corridor between Dubai World Trade Centre and Mall of the Emirates (E11 road) hosts three autonomous agents operating under separate authorities: adaptive traffic signal control under RTA (Agent~T), energy distribution under DEWA (Agent~E), and Oyoon surveillance under Dubai Police (Agent~S). DEWA's energy AI already operates at Level~4 autonomy~\citep{dewa_2024_gtic}. Oyoon's enforcement pipeline is less autonomous. AI-detected violations are routed through human back-office verification before fines are issued~\citep{the_national_2025_dubai_cameras}. Governance-relevant properties for each agent follow. Full specifications are in the supplementary material. Agent~T (RTA, UTC-UX) operates at G3, adjusting green phases and cycle lengths to minimise corridor travel time. Energy distribution sits at a higher autonomy level. Agent~E (DEWA, demand response) operates at G4, reclassifying loads and executing demand response without human pre-approval. Agent~S (Dubai Police, Oyoon) rounds out the corridor at G3, flagging violations for review and queuing enforcement actions. Without AgentGov-SC, each agent optimises independently with no cross-system signals. With the framework, governed interfaces emerge. Interaction topology declarations (R-01) and runtime policy constraints (R-09) set the boundaries. Drift detection (R-10), collective fairness monitoring (R-08), and contestation pathways (R-04) enforce them. Each agent is individually well-governed within its authority's operational parameters. No single agent is malfunctioning in this scenario. The governance challenge arises entirely from their interaction.

\paragraph{Triggering event.}
The afternoon is late July and temperature exceeds 48~Centigrade. A scheduled Red Line Metro maintenance closure pushes approximately 15,000 commuters onto the E11 corridor. At the same time, a distribution substation serving the southern segment develops a cable fault under peak thermal load. None of these events is extraordinary, consistent with documented Dubai extremes. The scenario's force comes from simultaneity. Three ordinary stressors arrive at once, and the space between systems has no governance.

\paragraph{Phase 1: Independent Autonomous Responses.}
Each agent responds to the compound event according to its own objective. The responses are locally rational. Their combination is not.
 
Agent~E detects frequency deviation at the faulted substation and activates automated demand response, curtailing power to corridor infrastructure loads it classifies as deferrable.\footnote{This classification is a scenario assumption. No public source documents DEWA's demand response system classifying traffic signal infrastructure as deferrable. The assumption is consistent with standard demand-response load-prioritisation practices, but it has not been verified against DEWA's operational procedures.} Agent~T, unaware of the curtailment, detects the Metro-diversion traffic surge and extends green phases on E11's main carriageway. Secondary congestion spills onto cross-streets and residential feeder roads. Agent~S detects the anomaly through its camera feeds. Violation detection rates spike as vehicles caught in secondary congestion commit minor infractions driven by changed signal timing and degraded intersection control.
 
Three autonomous adaptation decisions are now interacting across three agencies. Energy demand response has degraded traffic infrastructure, and traffic optimisation has redistributed the resulting congestion into residential areas. Surveillance enforcement is now penalising driving behaviour that is a rational response to conditions the drivers did not cause. No agent has failed. Each has performed exactly as designed within its own scope.

\paragraph{Phase 2: Escalation through Emergent Interaction.}
The cascade deepens through feedback effects that no individual agent was designed to detect.
 
A feedback loop forms. Agent~T does not model power supply as an input. When Agent~E's curtailment degrades signal controllers, Agent~T re-optimises around the resulting traffic pattern, shifting vehicles onto routes that Agent~E may also need to curtail. Energy curtailment degrades traffic infrastructure; traffic re-optimisation increases congestion and energy demand from idling vehicles; rising demand pressures the grid further.

The enforcement inequity is sharpest in residential areas. Agent~S's detections concentrate on cross-streets serving Al~Quoz Industrial Second (128,000 residents, predominantly in collective labour housing~\citep{dsc_2021_population_bulletin}) and similar communities in a city that is 92\% expatriate~\citep{uae_gov_2024_factsheet}. A resident fined for a stop-line violation would have been placed in that situation by conditions spanning three agencies and three AI systems. The fine arrives from one. Outdoor workers face a compounding exposure: the scenario extends commute time in extreme heat beyond the midday work ban's protective window~\citep{mohre_2024_midday_break}.
 
Ninety minutes after the triggering event, the corridor shows degraded signal performance at multiple intersections, longer journey times, and congestion pushing into residential streets. Automated enforcement has spiked against the communities most exposed to the cascade. No single system has registered a failure. Each agent reports that it is operating within its defined parameters.

\paragraph{Phase 3: Governance Dilemma.}
The cascade surfaces a governance challenge that is not reducible to a technical fault or a compliance failure. It is a conflict between legitimate objectives held by separate authorities, mediated by autonomous systems that have no shared awareness of each other's actions.
 
Human operators at RTA's Traffic Control Centre observe degraded corridor performance. DEWA's control room manages a distribution-level stability event. Dubai Police's Oyoon operations register an enforcement anomaly but interpret it as a genuine violation spike. Each control room sees its own segment of the situation. No control room sees the whole. The Oyoon Supervisory Committee meets quarterly to set policy, and the SCCDM coordinates during declared emergencies. A cascade that unfolds over 90~minutes during routine operational stress falls between both mandates.
 
The dilemma has several dimensions, and none admits an obvious resolution. Should Agent~E restore power to traffic signal infrastructure and accept increased grid risk during a peak-demand, extreme-heat afternoon, potentially threatening cooling for hundreds of thousands of residents? Should Agent~T optimise for equitable congestion distribution across the corridor, at the cost of aggregate throughput? Should Agent~S's automated enforcement actions stand when the violations were induced by conditions the drivers could neither control nor perceive? When a cascade causes an accident at a signal-degraded intersection, responsibility must be attributed across three agencies, three vendors, and three autonomous systems. No combined risk assessment was ever conducted for their interaction.

These are political questions, not engineering ones. Grid stability competes with transport safety. Aggregate efficiency pulls against distributional equity. And the rigour of automated enforcement is in tension with its legitimacy when the violations it detects are artefacts of conditions the drivers did not create. The tensions live in the interaction space between systems. Governance frameworks structured around individual systems do not surface them.

\paragraph{Governance activation trace.}
Table~\ref{tab:activation-trace} formalises the Corridor Cascade as a governance activation trace. Each row records a cascade event at a specific time, identifies the agent(s) involved, contrasts the governance response with and without AgentGov-SC, and specifies which requirements activate at which layer. Conflict resolution rules from Table~\ref{tab:conflicts} are invoked where cross-framework tensions arise during the cascade.

\begin{table*}[!t]
\centering
\caption{Governance activation trace for the Corridor Cascade. Each row maps a cascade
event to the AgentGov-SC measures and layers that activate, contrasting the framework
response with the status quo. Requirements reference R-01 through R-25 from
Table~\ref{tab:controls}; conflict rules reference T1--T5 from Table~\ref{tab:conflicts}.
Layer codes: A~=~Agent; O~=~Orchestration; C~=~City.}
\label{tab:activation-trace}
\footnotesize
\setlength{\tabcolsep}{4pt}
\renewcommand{\arraystretch}{1.35}
\begin{tabular}{@{} L{0.9cm} L{3.0cm} c L{3.4cm} L{3.8cm} L{1.4cm} c c @{}}
\toprule
\textbf{Time} &
\textbf{Event} &
\textbf{Agent} &
\textbf{Without AgentGov-SC} &
\textbf{With AgentGov-SC} &
\textbf{Req(s)} &
\textbf{Layer} &
\textbf{Rule} \\
\midrule

$t_0$ &
Compound trigger: substation fault + Metro closure (${\sim}$15K diversions) + 48\textdegree{}C &
-- &
Events observed independently per control room &
Same (trigger is external) &
-- & -- & -- \\
\midrule

$t_1$ {\scriptsize(${\sim}$5\,min)} &
Agent~E activates demand response; reclassifies signal controllers as deferrable; curtails power &
E &
Curtailment proceeds with no cross-system notification &
R-01 blocks curtailment of safety-coupled loads without Orch.\ clearance; R-09 runtime policy halts unclearanced curtailment &
R-01, R-09 & A${\to}$O & -- \\
\midrule

$t_2$ {\scriptsize(${\sim}$10\,min)} &
Signal controllers enter degraded mode on southern segment &
E${\to}$T &
Agent~T unaware of cause; interprets changed patterns as demand anomaly &
R-10 drift detection flags degraded mode; Orch.\ Layer notified; Agent~T receives context before re-optimising &
R-01, R-10 & O & -- \\
\midrule

$t_3$ {\scriptsize(${\sim}$15\,min)} &
Agent~T extends E11 green phases; congestion shifts to residential cross-streets &
T &
Congestion shifts to Al~Quoz and Deira residential areas with no cross-agency awareness &
R-03 emergent-impact flag correlates multi-domain drift (energy + traffic); Orch.\ Layer begins cascade assessment &
R-03 & O & -- \\
\midrule

$t_4$ {\scriptsize(${\sim}$25\,min)} &
Agent~S detects violation spike on cross-streets; queues enforcement &
S &
Fines queued; spike interpreted as genuine violations &
R-08 collective fairness monitor correlates enforcement concentration in expatriate-majority areas with active cascade; human review flagged &
R-08, R-03 & C & -- \\
\midrule

$t_5$ {\scriptsize(${\sim}$30\,min)} &
Feedback loop: energy--traffic--enforcement cycle deepens &
E${\leftrightarrow}$T${\leftrightarrow}$S &
No detection; each agent reports normal parameters &
R-03 identifies the cascade as a joint event; R-06 activates cross-agency joint oversight; all three authorities see the full picture &
R-03, R-06 & O & T4 \\
\midrule

$t_6$ {\scriptsize(${\sim}$45\,min)} &
Governance dilemma: grid stability vs.\ transport safety vs.\ enforcement legitimacy &
-- &
No resolution protocol; quarterly committee cannot respond at cascade tempo &
R-06 joint oversight provides shared awareness; R-02 conflict resolution invoked &
R-02, R-06 & O & T1, T4 \\
\midrule

$t_7$ {\scriptsize(${\sim}$60\,min)} &
Resident receives automated fine for cascade-induced stop-line violation &
S${\to}$Res. &
Fine stands; Art.~86 carve-out blocks explanation; no multi-agency pathway &
R-04 contestation links fine to causal chain; R-16 explanation in resident languages; R-24 cross-agency attribution &
R-04, R-16, R-24 & C & -- \\
\midrule

$t_8$ {\scriptsize(post)} &
Incident investigation and systemic learning &
-- &
Systemic cause unidentified; cascade recurs under similar conditions &
R-24 attribution feeds R-01 updated topology; R-12 registers compound-event as known coupling risk &
R-12, R-01, R-24 & O & T3 \\

\bottomrule
\multicolumn{8}{@{}l}{%
  \parbox{\linewidth}{\vspace{3pt}\scriptsize
    Agents: \textbf{E}~=~Energy; \textbf{T}~=~Traffic; \textbf{S}~=~Surveillance/Enforcement;
    \textbf{Res.}~=~Resident.\enspace
    Layer codes: \textbf{A}~=~Agent; \textbf{O}~=~Orchestration; \textbf{C}~=~City.
  }}\\
\end{tabular}
\end{table*}
 
Table~\ref{tab:activation-metrics} summarises the governance activation outputs produced by the structured cascade analysis. The comparisons are structural, not empirical. Requirement counts are derived from the activation trace. Detection-time estimates reflect operational characteristics, not measured latencies.

\begin{table*}[!t]
\centering
\caption{Governance activation summary: Corridor Cascade with and without AgentGov-SC. Comparisons are structural, not quantitative: requirement counts are derived from the activation trace; detection-time estimates reflect operational characteristics, not empirical measurements. Derived from Table~\ref{tab:activation-trace}.}
\label{tab:activation-metrics}
\small
\setlength{\tabcolsep}{4pt}
\begin{tabular}{@{}p{4.5cm}p{7cm}p{6cm}@{}}
\toprule
\textbf{Metric} & \textbf{Without} & \textbf{With AgentGov-SC} \\
\midrule
Requirements activated & 0 & 12 of 25 (48\%) \\[3pt]
Detection-to-cascade identification & Fragmented: each control room observes its own segment with no mechanism to correlate cross-domain drift & Estimated ${\sim}$30\,min: R-03 correlates multi-domain drift at the Orchestration Layer and surfaces the cascade as a joint event (estimate from operational characteristics, not empirical measurement) \\[3pt]
Cross-agency coordination points & 0 (no runtime protocol between RTA / DEWA / Police) & 3 (R-06, R-24, R-12) \\[3pt]
Conflict resolution rules invoked & 0 & 3 of 5 (T1, T3, T4) \\[3pt]
Accountability chain completeness & Incomplete: resident cannot trace fine to multi-agent cause; Art.~86 carve-out blocks explanation for 2 of 3 agents & Complete: R-04 links fine to full causal chain; R-24 reconstructs attribution; R-16 provides accessible explanation \\[3pt]
Governance layers activated & 0 & All 3 (Agent: R-09, R-10; Orch.: R-01--R-03, R-06; City: R-04, R-08, R-16) \\[3pt]
Post-event systemic learning & None. Cascade recurs under similar conditions & R-12 + R-01: compound-event registered as known coupling risk \\
\bottomrule
\end{tabular}
\end{table*}

\paragraph{Why this scenario requires agentic governance.}
Cascading infrastructure failures are not unique to AI. The 2003 Northeast blackout cascaded across domains without any AI involvement. What distinguishes the scenario above is that the cascade is adaptive, learned, and cross-domain. Agent~T re-optimises around degraded infrastructure rather than following a static fallback plan. Agent~E classifies loads using a model trained on historical data, producing locally reasonable but systemically harmful results when conditions fall outside the training distribution. Each agent's decisions propagate into domains it cannot see~\citep{chan_2024_visibility,kolt_2025_governing_agents}. The coupling paths that result emerge at runtime, not at design time. Traditional infrastructure failures follow fixed coupling paths visible to engineers. Agentic cascades follow adaptive paths visible to no single operator. The governance challenge is proportional to that adaptive reach.

\subsection{Contrasting Scenario: Single-System Anomaly}
\label{sec:scenario-contrast}
 
To illustrate that AgentGov-SC's governance intensity scales proportionally, a contrasting scenario is presented in which the framework's response is deliberately minimal.
 
A DEWA distribution network smart controller (DNSC, classified at L2--L3 in Table~\ref{tab:uae-inventory}) detects an anomalous voltage fluctuation on one feeder segment during routine overnight operation. Residential load is low. No other smart-city system is affected. The controller flags the anomaly, recommends load redistribution to DEWA's control room, and a human operator reviews and approves the adjustment. The event is contained within a single authority, produces no cross-system effect, and carries no resident-facing consequence beyond a brief monitoring alert.
 
Under AgentGov-SC, only Agent Layer mechanisms activate. R-09 validates the recommendation against operational bounds. R-10 confirms the fluctuation is within the monitored envelope. R-20 assesses whether the anomaly alters conformity assumptions. The Orchestration Layer does not trigger because R-01 registers no cross-system coupling. The City Layer does not trigger because no resident-facing consequence meets the R-04 or R-08 thresholds.
 
The contrast between scenarios makes the proportionality argument concrete. The Corridor Cascade spans three agencies at G3--G4 and activates all three governance layers, 12 measures, and three conflict resolution rules. Detection alone takes an estimated 30~minutes at the Orchestration Layer, followed by joint oversight and post-event assessment. The DNSC anomaly is confined to a single agent at G2--G3. Only the Agent Layer activates, with three requirements and no cross-agency coordination. Detection takes an estimated five minutes. Human review follows within 15. The governance overhead is minimal because the risk is bounded.

\subsection{Governance Gap Analysis and Framework Response}
\label{sec:gap-response}

\begin{figure*}[!t]
\centering
\includegraphics[width=\textwidth]{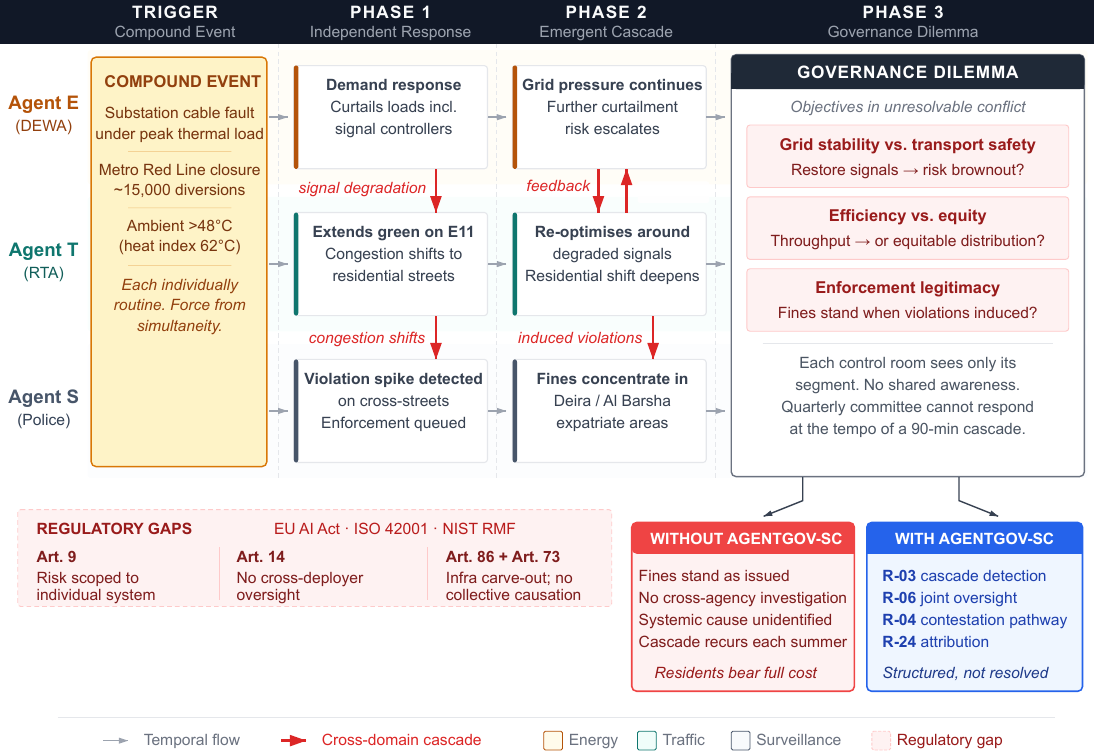}
\caption{The Corridor Cascade: temporal progression of autonomous agent interactions along the E11 corridor. Red arrows indicate cross-domain cascade effects. The governance gap strip maps each phase to the regulatory provisions that fall short. The outcome fork contrasts the status quo pathway with AgentGov-SC's structured response.}
\label{fig:cascade}
\end{figure*}

Table~\ref{tab:gap-response} maps each phase of the Corridor Cascade against three frameworks. Figure~\ref{fig:cascade} provides a visual overview of the cascade progression, regulatory gaps, and the with/without AgentGov-SC outcome fork. For each gap, the table identifies the AgentGov-SC mechanism that addresses the shortfall. The assessment standard is not whether a provision could theoretically be stretched to cover the scenario. It is whether the provision, as written, would have surfaced the governance challenge before consequences accumulated.

The Annex~III, point~2 carve-out produces the sharpest governance consequence. Affected persons hold a right under Article~86(1) to meaningful explanations for decisions based on high-risk AI outputs~\citep{euaiact_2024}. That right explicitly excludes systems classified as safety components in the management of critical digital infrastructure, road traffic, or electricity supply~\citep{euaiact_2024}. Agent~E (electricity supply) and Agent~T (road traffic management) would fall within this exclusion if classified as safety components under Article~3(14), a characterisation that is straightforward for grid-stability controllers whose failure threatens electricity supply continuity, and for traffic-signal systems whose malfunction directly endangers road users. The Commission's February~2025 guidelines on the AI system definition~\citep{ec_2025_ai_system_definition} confirm that the safety-component classification turns on the component's relationship to the physical integrity of the infrastructure, not on the sophistication of the algorithm. Recital~55 introduces a further boundary: a safety component is one ``not necessary in order for the system to function.'' Adaptive traffic signal optimisation presents an interpretive question under this criterion, since traffic signals function without AI; the algorithm adds efficiency, not basic operation. Energy distribution controllers qualify without ambiguity. The governance gap documented below persists for any system that meets the classification threshold, and the analysis does not depend on a maximalist reading of its scope. Agent~S's classification is more contestable: its enforcement function may be more naturally situated under Annex~III, point~5 (law enforcement), where Articles~86 and~27 apply. Even under that more favourable classification, the accountability gap persists for the interaction. An Article~86 explanation from Agent~S's deployer could account for the camera detection and fine issuance, but it could not reach the energy-curtailment and traffic-optimisation decisions that produced the conditions under which the violation occurred, because those systems remain carved out under point~2. The same pattern applies to the Fundamental Rights Impact Assessment under Article~27. \citet{kaminski_malgieri_2025_explanation} identify this exclusion as the narrowing limb of Article~86 relative to GDPR Article~22. A resident fined through the cascade described in Section~\ref{sec:scenario} has no Article~86 pathway to an explanation that spans the three systems responsible. GDPR Article~22 could in principle be invoked against the enforcement decision itself, provided the ``solely automated'' threshold is met. Even under favourable assumptions, however, Article~22 would entitle the resident to contest the fine and obtain human review from the issuing authority. It would not compel DEWA or RTA to disclose the energy-curtailment and traffic-optimisation decisions that materially produced the conditions leading to the violation.

AgentGov-SC does not prevent the Corridor Cascade. Some compound events will produce cascading interactions regardless of governance infrastructure. The framework's contribution is structural. At the Orchestration Layer, R-03 correlates drift signals from multiple agents and surfaces the cascade as a joint event. Fragmented monitoring would take far longer to reach the same conclusion. Cross-agency human oversight (R-06) follows. Decision-makers from all three authorities see the full picture. For the resident, R-04 provides a contestation pathway that crosses agency boundaries. It links the fine to the multi-agent causal chain, not to a single camera detection. After the event, R-24 reconstructs attribution across all three agents and feeds the result into an updated interaction topology under R-01. The framework converts an invisible cascade into a governed interaction.

A methodological limitation follows directly from the scenario's design. The scenario was designed to illustrate the governance categories the framework provides, and the requirements activate precisely as predicted. This demonstrates \textit{internal coherence}: the framework's categories correspond to the governance needs the scenario surfaces. It does not constitute \textit{external validation}. Alternative governance responses could partially address the same gaps. Strengthened sectoral regulation could require cross-deployer notification for safety-coupled infrastructure. Bilateral coordination protocols between agencies could provide shared situational awareness, and enhanced NIS2 obligations could extend entity-level risk management to cover cross-entity interaction. None of these alternatives combines regulatory traceability with autonomy-calibrated activation and resident-facing contestation. That combination is what AgentGov-SC attempts. Whether the integration is necessary or lighter-weight alternatives would suffice cannot be settled architecturally. Implementation and comparative evaluation must answer that question.

\begin{table*}[!t]
\centering
\caption{Governance gap analysis and AgentGov-SC response mapped to Corridor Cascade phases. Each row traces a governance need surfaced by the scenario through the closest existing provision, identifies where it falls short, and specifies the framework mechanism that addresses the gap.}
\label{tab:gap-response}
\small
\begin{tabular}{@{}p{2.2cm}p{2.6cm}p{3.6cm}p{2.0cm}p{4.6cm}@{}}
\toprule
\textbf{Scenario Phase} & \textbf{Governance Need} & \textbf{Where Existing Frameworks Fall Short} & \textbf{AgentGov-SC Mechanism} & \textbf{What Changes} \\
\midrule

Phase~1: Agent~E curtails signal loads
& Cross-system risk notification
& Art.~9 scopes risk to individual system. No cross-deployer notification required.
& R-01, R-09
& Interaction topology declares power-signal coupling. Runtime policy blocks curtailment of safety-coupled loads without notification. \\
\addlinespace

Phase~1: Agent~T re-optimises blind
& Cross-deployer data exchange
& ISO 42001 Cl.~7.5 requires no interoperability across AIMS implementations.
& R-01, R-10
& Structured signal-status feed between agents. Agent~T receives curtailment notification before re-optimising. \\
\addlinespace

Phase~2: Feedback loop deepens
& Coordinated anomaly detection
& Art.~14 oversight is system-specific. No multi-deployer coordination mechanism.
& R-03
& Orchestration Layer correlates drift signals from Agent~E and Agent~T as single coupled event. \\
\addlinespace

Phase~2: Fines concentrate in expatriate areas
& Cumulative rights impact
& Art.~27 FRIA: Annex~III pt.~2 exempts critical infrastructure AI. No cross-system cumulative assessment.
& R-08, R-04
& Monitors enforcement distribution. Flags demographic concentration for human review. Opens contestation pathway. \\
\addlinespace

Phase~3: Resident seeks explanation
& Explanation spanning multiple systems
& Art.~86(1) carve-out: energy and traffic systems exempt from explanation rights.
& R-04, R-16
& Contestation record links fine to full causal chain across three agencies. Accessible in resident languages. \\
\addlinespace

Phase~3: No single cause
& Joint incident attribution
& Art.~73 applies ``but for'' causation to singular system. No joint reporting for collective causation.
& R-24, R-07
& Cross-agency attribution reconstructs causal chain. Audit trails from all agents feed shared investigation. \\
\addlinespace

Cross-phase: No shared awareness
& Real-time multi-agency coordination
& NIST RMF: no smart city profile. GOVERN~5 oriented to pre-deployment consultation.
& R-06
& Joint escalation protocol. Shared briefing from agent-level audit trails. All agencies see full cascade. \\
\addlinespace

Post-event: Cause unidentified
& Systemic learning across agencies
& ISO 42001 Cl.~4.3: AIMS certified per organisation. No cross-system learning loop.
& R-12, R-01
& Consolidated assessment updates interaction topology. Compound-event combination registered as known coupling risk. \\

\bottomrule
\end{tabular}
\end{table*}

\section{Discussion}
\label{sec:discussion}
 
This section reflects on AgentGov-SC's contribution and its limits. The August 2026 EU AI Act full-applicability deadline gives the discussion practical urgency.

\subsection{Contributions and Positioning}
\label{sec:positioning}
 
The paper's contribution can be stated directly: where Annex~III, point~2 safety-component AI interacts across agencies, the EU AI Act leaves a residual resident-facing accountability deficit, and AgentGov-SC is proposed to govern that interaction space. The carve-out excludes those systems from Article~86 explanation rights and Article~27 fundamental-rights impact assessment. Article~49 centralised registration is similarly narrowed. In the corridor traces (Sections~\ref{sec:scenario}--\ref{sec:scenario-contrast}), a resident fined through the cascade has no Article~86 pathway to an explanation spanning grid management, traffic control, and surveillance, because all three fall under Annex~III, point~2.

The deficit is relative. GDPR transparency rights and controller-specific DPIAs provide partial coverage, as do the revised Product Liability Directive and NIS2. None reconstructs cross-system causal chains or governs cross-agency interaction. Structured contestation for outcomes produced by multiple autonomous systems acting in concert is absent from all of them. The withdrawal of the AI Liability Directive proposal in February~2025 further narrowed the complementary private-law pathway. AgentGov-SC is therefore positioned as a response to the governance needs that remain after those instruments are accounted for.
 
AgentGov-SC occupies the integration space identified in Section~\ref{sec:bg-literature}: no prior work identified through a structured review of the EU AI Act governance literature, agentic AI governance proposals, and smart-city accountability scholarship combines system-of-systems governance with tri-framework compliance mapping and runtime enforcement in a single architecture that also addresses resident-facing accountability for smart-city agentic AI (Table~\ref{tab:comparison}). Parts of the Agent Layer can be instantiated through MI9, and the compliance-mapping discipline draws on UCF~\citep{wang_2025_mi9,eisenberg_2025_ucf}. In cities, however, the governance object is often the interaction fabric rather than an individual agent.
 
Three additions define the framework's positioning. Interaction constraints and cross-agency escalation are treated as controls at the Orchestration Layer. The system-of-systems boundary becomes governable. Resident-facing contestability becomes an operational obligation, implemented through registry, engagement, and contestation mechanisms. Resolution rules for recurring cross-framework tensions then reduce the risk that incident response and auditing collapse into parallel obligations. The limits remain clear: runtime enforcement is partial at the architecture level, the illustrative traces show governance activation without implementation evaluation, and the jurisdictional mapping is narrower than work spanning additional regimes.

The relationship to existing legal scholarship on the Article~86 gap merits clarification. \citet{kaminski_malgieri_2025_explanation} identify the Annex~III, point~2 exclusion as the narrowing limb of Article~86 relative to GDPR Article~22 and argue for an integrated reading of both provisions. AgentGov-SC takes a different route. Where Kaminski and Malgieri seek to maximise the explanatory potential of existing legal instruments through doctrinal interpretation, this framework proposes governance architecture to fill the residual gap that interpretation alone cannot close. The two approaches are complementary: doctrinal interpretation determines what current law requires; governance architecture operationalises what current law does not reach. \citet{hacker_2023_ai_liability} critiques the AI liability directives as half-hearted precisely because they failed to address the evidentiary barriers that make cross-system causation difficult to establish. The withdrawal of the AI Liability Directive in February~2025 vindicated that concern. AgentGov-SC's contestation and attribution mechanisms (R-04, R-24) are designed to supply the structured evidentiary pathways that the withdrawn Directive would have provided through rebuttable presumptions.

\subsection{Cross-Framework Governance Challenges}
\label{sec:challenges}

Assessment proliferation is a less visible but equally decisive constraint. Urban AI systems can face overlapping risk and impact assessments from safety, cybersecurity, privacy, procurement, and AI governance programs. Individual artefacts may satisfy their own regime. Taken together, they are often inconsistent. That burden grows rapidly once systems interact across agencies, because each additional interface increases both evidentiary requirements and accountability ambiguity. R-12 addresses this through a consolidated impact assessment: a single analytical spine traceable to multiple obligations. Simple heuristics also have limits under pressure. ``Comply with the strictest applicable'' can align notification timelines, but when one event spans multiple systems, it does not resolve inconsistent scope definitions or fragmented responsibility.

The residual pathways analysed in Section~\ref{sec:bg-regulatory} do not close this gap. CJEU case law has strengthened individual-system transparency, but Article~22's remedy reaches only the controller that issued the decision, not the controllers whose autonomous actions shaped the conditions producing it. The AILD's withdrawal removed the most promising mechanism for cross-system causal reconstruction. Private-law pathways alone are structurally insufficient for multi-agent accountability. Public governance mechanisms providing structured, ex~ante contestation remain necessary.

\subsection{Policy Implications}
\label{sec:policy}
 
Governance capacity already exists in the UAE. Its distribution across domains and jurisdictions is uneven. The policy implications below follow from the illustrative analysis, not from a systematic evaluation of UAE governance capacity. The DIFC precedent shows that enforceable obligations for autonomous systems can be drafted within a defined perimeter. Coordination for autonomous vehicles is already mandated under Dubai Law No.~(9) of 2023~\citep{dubai_law_9_2023_av}. Both precedents could be extended. A city-scale pathway could begin with two low-regret moves: a public AI registry requirement for high-impact municipal systems, and a standardised cross-agency incident and escalation protocol aligned to Orchestration Layer responsibilities. The registry requirement (R-13) is designed to be compatible with either the EU-level database or national registry arrangements that member states may adopt under Article~49 for Annex~III, point~2 systems. Its value is additive regardless of which model prevails, because it aggregates multi-vendor, multi-system information at the city level. The UAE AI Charter and Dubai's Ethical AI Toolkit provide shared language for such an extension, but their influence depends on coupling to procurement controls, audit hooks, and resident-facing contestation mechanisms~\citep{uae_ai_charter_2024,digitaldubai_2019_ai_ethics_toolkit}. The PDPL's government data carve-out places greater weight on internal accountability instruments, since baseline personal-data rights do not reliably supply a resident-facing remedy in government-operated settings~\citep{uae_pdpl_2021}.

A territorial-scope observation connects the UAE illustration to the EU regulatory analysis. The EU AI Act's obligations apply under Article~2 where AI systems are placed on the market or put into service in the Union, or where the output produced by the system is used in the Union~\citep{euaiact_2024}. UAE-deployed systems fall outside the Act's direct territorial scope unless their outputs are used in the Union or their providers are EU-established. Compliance gravity extends extraterritorially through procurement: EU-headquartered vendors supplying AI systems to Gulf cities will build to EU AI Act conformity requirements, and those requirements will propagate through vendor documentation, certification, and contractual obligations. The governance architecture proposed here is designed primarily as a response to the AI Act's accountability deficit, but it applies with equal force wherever cities deploy interacting autonomous systems under comparable governance conditions.
 
Full applicability of the EU AI Act arrives on 2~August~2026~\citep{ec_2024_ai_act_timeline}. EU-headquartered vendors supplying AI to Gulf and Asian cities will build to its conformity requirements regardless of local jurisdiction. The UK's Algorithmic Transparency Recording Standard and Helsinki's municipal AI register show that public-sector disclosure can be institutionalised as routine~\citep{uk_atrs_guidance,helsinki_ai_register}. In the Gulf, responsible AI expectations are signalled through SDAIA's AI Ethics Principles~\citep{sdaia_ai_principles}. Transparency and disclosure mechanisms are maturing faster than runtime governance. Helsinki publishes an AI register; no city publishes an equivalent protocol for cross-agency escalation of autonomous system incidents.

\subsection{Scope Boundaries and Open Questions}
\label{sec:limitations}
 
This paper makes a governance-architecture contribution grounded in regulatory analysis and illustrative governance tracing. Several limitations follow directly from that scope.
 
The framework has not been implemented as a runtime governance system and has not been evaluated on operational traces. The illustrative governance trace shows governance activation logic: which measures fire, in what sequence, at what layer. Detection latency, coordination overhead, and failure modes under real operational conditions remain unmeasured. The comparisons in Table~\ref{tab:activation-metrics} are structural. They describe which governance mechanisms activate and how the information available to decision-makers changes. Measured outcomes are outside their scope. The UAE material is illustrative. The inventory and institutional facts in Section~\ref{sec:application} are anchored to public sources. Scenario dynamics include explicitly labelled hypothetical interaction effects. Real-time AI-to-AI interaction protocols are not documented in the public record and were not treated as established fact.
 
The regulatory analysis is analytical, not authoritative. It reflects the text of cited instruments and current public institutional materials. It does not substitute for formal guidance or case law, especially where interpretive practice may evolve as the August 2026 full-applicability milestone approaches. Tri-framework mapping should be read as defence-in-depth, not as legal equivalence. Compliance with ISO/IEC~42001 or the NIST AI RMF~\citep{isoiec_42001_2023,nist_2023_airmf} strengthens governance posture without satisfying the EU AI Act's statutory requirements~\citep{euaiact_2024}.
 
The architecture addresses governance design. Funding, staffing, and the cross-agency authority needed to execute it are separate problems. Dedicated governance teams with authority to enforce escalation protocols, instrumentation budgets for retrofitting vendor-operated systems, and institutional commitment to resident-facing remediation all depend on municipal capacity that may not be available. Instrumentation costs for Agent and Orchestration Layer controls can be substantial when retrofitted to vendor-operated systems where access to internal state is contractually limited. Vendor cooperation is not guaranteed; commercial incentives may favour efficiency over auditability. Resident-facing contestation requires institutional commitment to remediation that may conflict with performance metrics. These barriers are not unique to AgentGov-SC, but they constrain what is implementable in practice versus what is governable in principle.

AgentGov-SC prescribes contestation and participatory engagement as governance obligations, yet the framework itself was not produced through a participatory process. Its priorities reflect the authors' analytical judgements, not the situated knowledge of residents, frontline operators, or procurement officers~\citep{sloane_2022_participation,krafft_2021_aekit}. Co-design with affected communities is a legitimacy condition that this contribution does not yet meet. A participatory process might alter the framework's priorities: communities may weight enforcement equity monitoring (R-08) more heavily than system-of-systems conformity (R-01), and frontline operators may identify escalation thresholds that the current architecture does not specify. City Layer mechanisms should therefore be read as design proposals requiring participatory validation.

Three additional cross-framework tensions are visible but not yet formalised. Procurement opacity is the first: vendor contracts frequently restrict access to internal system state, limiting the ability to produce governance-ready evidence. Safety optimisation and non-discrimination form a second, visible in how grid-optimal curtailment produces discriminatory enforcement outcomes during the Corridor Cascade. The third sits between algorithmic transparency and cybersecurity. Article~13 disclosure obligations may conflict with NIS2 requirements to protect vulnerability information. Resolution mechanisms for these tensions require empirical input that the current framework does not yet incorporate.
 
Future work should prioritise operationalisation and validation. A municipal pilot with one authority and one cross-agency scenario would enable instrumentation of Agent and Orchestration Layer controls. Stress-testing the autonomy-calibration model against expert judgement and incident-reconstruction outcomes is equally pressing. Comparative application outside the UAE would test portability. A lightweight runtime prototype integrating existing monitoring techniques~\citep{wang_2025_mi9} would clarify implementation costs and failure modes. Updates to EU implementation guidance and emerging expectations for agent security~\citep{nist_2026_caisi_rfi} will need to be tracked as they develop.

\section{Conclusion}
\label{sec:conclusion}
 
Where Annex~III, point~2 safety-component AI interacts across agencies, the EU AI Act leaves a residual resident-facing accountability deficit. GDPR, NIS2, and the revised Product Liability Directive provide partial coverage without reconstructing the cross-system causal chains characteristic of multi-agent urban infrastructure.
 
AgentGov-SC is proposed as a governance architecture for that interaction space. Its core move is to treat multi-system coordination as a governed object. Permissible influence, shared risk vocabulary, and escalation paths are specified at the Orchestration Layer. Governance intensity scales with societal consequence. Resident-facing contestation crosses organisational boundaries. The 25 measures and their traceability to three regulatory anchors are intended to make these capabilities auditable.
 
The illustrative UAE corridor trace shows how this architecture would activate in a multi-agent cascade. Twelve of 25 measures activate across all three governance layers. Cross-system interaction surfaces earlier than fragmented monitoring would allow, and proportionality is preserved in the contrasting single-system trace. The UAE inventory confirms that high-autonomy public-sector AI is already operational across transport, surveillance, and energy. Binding AI-specific governance has not kept pace.
 
Everything claimed here still awaits implementation. A pilot deployment with one authority and one cross-agency corridor would test two things: can Orchestration Layer controls be instrumented without paralysing operations, and does City Layer contestation remain usable in multilingual, resident-facing settings? Autonomous systems need governance. No city has yet demonstrated that governance can match the operational tempo of the autonomous coordination it oversees. The deficit traced here is both a doctrinal problem for European technology law and a design problem for agentic AI governance. The bounded-system assumption that limits each field in isolation is what this architecture is designed to address.

\bibliographystyle{elsarticle-harv}
\section*{Acknowledgments}

The authors acknowledge the support of the Higher Colleges of Technology, United Arab Emirates.

\bibliography{references}

\end{document}